\documentclass[preprint,12pt]{elsarticle}

 \usepackage{graphicx}
 
 \usepackage{float}
 
  \usepackage{algorithm}
 \usepackage{algorithmic}

\usepackage{color}
\usepackage{hhline}
\usepackage{mathrsfs}
\usepackage{graphicx}
\usepackage{dcolumn}
\usepackage{bm}
\usepackage{multirow}
\usepackage{booktabs}
\usepackage{afterpage}
\usepackage{amsmath}
\usepackage{ulem}

\newcommand{\algdo}{\algorithmicdo}
\newcommand{\algend}{\algorithmicend}

\newcounter{bla}

\journal{Computer Physics Communications}

\begin{document}

\begin{frontmatter}




\title{Non-empirical weighted Langevin mechanics for the potential escape problem: parallel algorithm and application to the Argon clusters}

\author[a]{Yuri S. Nagornov\corref{author}}
\author[a]{Ryosuke Akashi}

\cortext[author] {Corresponding author.\\\textit{E-mail address:} iurii@cms.phys.s.u-tokyo.ac.jp}
\address[a]{Department of Physics, The University of Tokyo, Hongo, Bunkyo-ku, Tokyo 113-0033, Japan}

\begin{abstract}
Recently a non-empirical stochastic walker algorithm has been developed to search for the minimum-energy escape paths (MEP) from the minima of the potential surface [J. Phys. Soc. Jpn. 87, 063801 (2018)]. This algorithm is novel in that it tracks the MEP monotonically and does not use the whole Hessian matrix but only gradient and Laplacian of the potential. In this work, we implement an MPI-parallelized version of this algorithm in a simple way. We also explore efficient ways to reduce the number of walkers required for the accurate tracking of the MEP and generate initial positions automatically. We apply the whole scheme to the Lennard-Jones argon cluster with 7-38 atoms to demonstrate the successful tracking of the reaction paths. This achievement paves the path to non-empirical simulation of rare reactions without coarse-graining or artificial potential. 
\end{abstract}


\end{frontmatter}

\newpage






\section{Introduction}

Efficient simulation of structural transitions and chemical reactions of matter in atomic scale--which concerns folding of proteins, molecular reactions, diffusion of impurities in solids, nucleation, etc.--is a common issue across a wide range of research fields - chemistry, physics, biology and their intersections. The difficulty of this task originates from the rarity of these phenomena in the simulation time scale; the time step $\Delta t$ for the simulation is required to be as small as picoseconds or even sub-femtoseconds, whereas the typical timescale of the phenomena is more than microseconds. This difference in the time scales renders their straightforward simulations almost impossible and alternative ways are usually sought. This paper concerns a development of the method to manage the simulation of a class of the rare events described by infrequent transitions between the minima of the potential surface (or free-energy surface) defined in the configuration space, induced by thermal fluctuation.

The frequency of the transition across the potential minima under the thermal fluctuation is dominated by the minimum potential energy barrier between them, and therefore the seeking of the paths connecting the potential minima with minimum energy barrier (minimum energy paths; MEPs) is of particular importance. Many existing methods can be applied when either the product (destination of the transition in the configuration space) or the reaction coordinate (collective coordinates which well characterize the transition) is given {\it a priori}, for example, the blue-moon~\cite{Carter-Ciccotti-bluemoon,Sprik-Ciccotti-bluemoon}, targeted dynamics~\cite{Schlitter-Wollmer-targetedMD}, steered dynamics~\cite{SteeredMD-orig-Sci1996} hyperdynamics~\cite{Voter-hyperdyn-JCP1997,Voter-hyperdyn-PRL1997}, metadynamics~\cite{Laio01102002}, adaptive biasing force~\cite{Darve-ABF-JCP2001} and many others. In the former case, various path optimization or sampling methods are available, whereas in the latter, one can add artificial force with respect to the reaction coordinates to the potential force ~\cite{Maeda-Morokuma-AFIR2010} or execute a weighted sampling dynamics referring to the coordinates~\cite{Allen-tenWolde-FFS-PRL2005,Harada-Kitao-PaCSMD-JCP2013,Harada-Shigeta-SDS-JCTC2017,Harada-Shigeta-SDS-JCC2017}.

Recently, the authors has proposed an atomistic simulation method that realize the transition between the potential minima by tracking the path from a minimum to the saddle point through the valley of the potential surface~\cite{Akashi}. This method does rely neither on the a priori knowledge of the product nor the reaction coordinates. The goal of this development is to establish a non-empirical simulation method that is applicable to the situation when the possible products are unknown. Although there are also methods aiming at the same goal such as the eigenvector following~\cite{Hilderbrandt-NewtonRaphson, Cerjan-Miller-eigenvec-follow}, gentlest ascent~\cite{gentlest-ascent-Zhou2011} and dimer methods~\cite{Dimer1999}, the present method has advantages and disadvantages: It does not require us to calculate the Hessian matrix of the potential and its eigenpairs and yields monotonic tracking of the MEPs. Instead, we need to run a number of walkers interacting through a stochastic process. Further development of the method could therefore give us an interesting alternative for the abovementioned goal. 

The method has been demonstrated only with a very simple two-dimensional potential surface. In this study, we seek for the possible application to larger systems. In the path search in the higher dimensional systems, the entire volume of the configuration space diverges, which is why the structure and path search is generally difficult. Nevertheless, the dimension of the MEPs is in principle always one and the number of the relevant paths is, in the low temperature, a few. With these contrasting facts we conceive that the path search that tracks the valley lines to the saddle points could be possible with computational cost and time that slowly increase with respect to the dimension of the system.

\section{Non-empirical weighted Langevin mechanics for tracking minimum energy escape paths}

Here in a brief we show concept, mathematical model and then main algorithmic steps of our method published in Ref.~\cite{Akashi}. We address the molecular mechanics described by the following Langevin equation
\begin{eqnarray}
dx_{i} =  -\frac{\partial_{i} U({\bm x})}{\Gamma} dt+ \sqrt{\frac{2k_{\rm B} Tdt}{\Gamma}}W_{i}
,
\label{eq:SDE-x}
\end{eqnarray}
where $U({\bf x})$, $k_B$ and $T$ are the potential, Boltzmann constant and temperature, respectively.
$i$ is the index for the degrees of freedom.
${\bf W}$ is the vector whose components are randomly generated from the standard normal distribution at each step. 
$\Gamma$ is the friction constant.
This equation represents the motion of the degrees of freedom ${\bf x}=\{x_{1}, x_{2}, \dots\}$ under the potential $U({\bf x})$ and thermal fluctuation. When the temperature $T$ is low, the trajectories escaping from one minimum is observed with extremely low probability. In order to see how such trajectories can be rendered more frequent, we first move to the Smoluchowski equation describing the distribution function of ${\bf x}$, $p({\bf x}, t)$, given by~\cite{Gardiner-book}
\begin{eqnarray}
\partial_t p({\bm x},t) &=& \frac{1}{\Gamma}\partial_i [ (\partial_{i}U({\bm x})) + k_{B}T \partial_i]p({\bm x},t)
\\
&\equiv& \hat{L}_{\rm Sm}p({\bm x},t)
.
\label{eq:Smoluchowski}
\end{eqnarray}
This equation has a general intriguing property that, regardless of the initial distribution $p({\bm x},t=0)$, it converges to the Boltzmann distribution $p_{\rm B}({\bm x})$ in the infinite $t$ limit. This means that, starting from any localized initial distribution $p({\bm x},t=0)~\delta({\bm x}-{\bm x}_{0})$, there must be a component in $p({\bm x}, t)$ that climbs up the potential slope. An intuitive way to enhance such component is to divide it by $p_{\rm B}({\bm x})$.

Along this line of consideration, the property of the modified distribution $q({\bm x}, t)\propto p({\bm x}, t) / [{\rm exp}(-(1-\delta)U({\bf x})/(k_{\rm B}T))]$ has been examined with one-dimensional harmonic potential $U(x)=\alpha x^2/2$ \cite{Akashi}. The analytical solution of $q({\bm x}, t)\propto p({\bm x}, t)$ with the initial condition $p({\bm x}, 0)=\delta({\bm x}-{\bm x}_0)$ has been found to have a property as follows: (i) It has the Gaussian form and its width gradually spreads with $t$, (ii) if $\delta<1/2$, its center first increases by $\propto x_{0}t$ and, as expected from the stationary solution $p_{\rm B}({\bm x})$, converges to the center of the parabola in a long time. A scenario is conceived from these features: In many-dimensional space, If ${\bm x}_{0}$ is placed along the ``valley line" of the potential, where in one direction the potential has nonzero gradient and in others zero, the center of $q({\bm x}, t)$ goes farther along the line upward for a while. By tracking the center of $q({\bm x}, t)$, we can therefore draw the MEPs as the valley lines extending from a potential minimum. This simplistic argument has been actually validated in the two-dimensional case~\cite{Akashi} with the algorithm appended later.



We then construct a microscopic stochastic algorithm to reproduce $q({\bm x},t)$. From the Smoluchowski equation [Eq.~(\ref{eq:Smoluchowski})], the corresponding Master equation for $q({\bm x},t)$ with general transformation $p({\bm x},t)=C(t){\rm exp}[-V({\bm x})/k_{\rm B}T]q({\bm x},t)$ is given by~\cite{Giardina-review-JStatPhys2011, Akashi} 
\begin{eqnarray}
\partial_{t}q({\bm x},t)=\{ \hat{L}'_{\rm Sm} + \hat{L}_{\rm rate}\}q({\bm x},t)
\label{eq:Master-eq}
\end{eqnarray}
with
\begin{eqnarray}
\hat{L}'_{\rm Sm}
&=&
\frac{1}{\Gamma}
\partial_{i} [\partial_i (U({\bm x})-2V({\bm x}))]+\frac{k_{\rm B}T}{\Gamma}\partial_{i}^2,
\\
\hat{L}_{\rm rate}
\equiv
L_{\rm rate}({\bm x}, t)
&=&
\frac{1}{\Gamma}
\left[
F({\bm x})-\langle F\rangle_{q(t)}
\right],
\label{eq:L-rate-def}
\\
\partial_t {\rm ln}C(t) &=& \frac{1}{\Gamma}\langle F\rangle_{q(t)},
\label{eq:C-timeevol}
\\
F({\bm x})
&=&
\partial_{i}^{2}V({\bm x})
+
\frac{1}{k_{\rm B}T}(\partial_{i}V({\bm x}))[\partial_{i}(V({\bm x})-U({\bm x}))]
.
\label{eq:Ffunc}
\end{eqnarray}
Coefficient $C(t)$ is introduced so that the norm of $q$ is kept to unity. We define the average of the function $f({\bm x})$ by $\langle f\rangle_{q(t)} = \int d{\bm x} q({\bm x},t)f({\bm x})$. The finite time evolution of $q$ is formally represented as
\begin{eqnarray}
q({\bm x},t+\tau)
={\rm exp}\{[\hat{L}'_{\rm Sm}+\hat{L}_{\rm rate}] \tau\}q({\bm x},t)
.
\label{eq:q-evolve}
\end{eqnarray}

We then construct the stochastic process that operates on the walkers having individual values of ${\bm x}$ and reproduces $q({\bm x}, t)$ as their aseembly. The Suzuki-Trotter decomposition~\cite{Suzuki-Trotter-1-AMS1959,Suzuki-Trotter-2-CMP1976,note-Trotter} of Eq.~(\ref{eq:Master-eq}) yields:
\begin{eqnarray}
&&{\rm exp}\{[\hat{L}'_{\rm Sm}+\hat{L}_{\rm rate}] \tau\}
\nonumber \\
&&\simeq
{\rm exp}\{\hat{L}'_{\rm Sm} \tau/2\}
{\rm exp}\{\hat{L}_{\rm rate} \tau\}
{\rm exp}\{\hat{L}'_{\rm Sm} \tau/2\}
.
\label{eq:timestep-decompose}
\end{eqnarray}

In terms of the walker dynamics, operator ${\rm exp}\{\hat{L}'_{\rm Sm} \tau/2 \}$ is the time evolution
\begin{eqnarray}
&&{\bm x}(t+\tau)-{\bm x}(t) \nonumber \\
&&=-\frac{1}{\Gamma}\nabla (U({\bm x}(t))-2V({\bm x}(t)))\tau +  \sqrt{\frac{2k_{\rm B}T\tau}{\Gamma}} {\bm W}
\label{eq:Lang_step}
\end{eqnarray}
applied to all the walkers. Operator ${\rm exp}\{\hat{L}_{\rm rate} \tau \}$, being simply a multiplication of a function of ${\bm x}$ on $q({\bm x}, t)$ to $q({\bm x}, t)$, is implemented as the importance sampling~\cite{Grassberger-importance-sampling-PRE1997} as described below. The statistical average of physical quantity $f({\bm x})$ with respect to $q({\bm x}, t)$, which is used in Eqs.~(\ref{eq:L-rate-def}) and (\ref{eq:C-timeevol}), is then recast to the average with respect to the walkers 
\begin{eqnarray}
\hspace{-10pt}
\langle f({\bm x})\rangle_{q(t)}
\equiv
\int d{\bm x} f({\bm x}) q({\bm x}, t)
\simeq
\frac{1}{N_{\rm w}}\sum_{i=1}^{N_{\rm w}} f({\bm x}_{i}(t))
.
\end{eqnarray}

We here append the specific procedure for the time evolution Eq.~(\ref{eq:timestep-decompose}) for each time step:
\begin{enumerate}

\item
${\rm exp}\{\hat{L}'_{\rm Sm} \tau /2\}$:
\\
Evolve the walkers independently by the Langevin equation Eq.~(\ref{eq:Lang_step})

\item
${\rm exp}\{\hat{L}_{\rm rate} \tau\}$: \\
Do the following for all the walkers $i=1$--$N_{\rm w}$ independently. If $r \equiv {\rm exp}\{L_{\rm rate}({\bm x}_{i}(t), t)\tau\}<1$, remove that walker by probability $1-r$; otherwise, add $\lfloor (r-1) \rfloor$ walker(s) with the same value of ${\bm x}$ as $i$ and add one more by probability $r-1 -\lfloor (r-1) \rfloor$. Modify $N_{\rm w}$ accordingly.

\item
${\rm exp}\{\hat{L}'_{\rm Sm} \tau/2\}$: \\
Repeat step (1)

\end{enumerate}

According to the argument above, we have proposed as a specific form of $V({\bm x})$ as 
\begin{eqnarray}
V({\bm x})=(1-\delta)U({\bm x}).
\label{eq:V-def} 
\end{eqnarray}
With this form, the original potential function $U({\bm x})$ is straightforwardly devised and its convex/concave structure is reflected. Note that the Laplacian of $U({\bm x})$ is then needed to evaluate $L_{\rm rate}({\bm x})$ for each walker. We have demonstrated that this setting enables us to track the valley lines of $U({\bm x})$ from a known potential minimum to the saddle points for a case of two-dimensional potential surface. In this work, we develop systematic algorithms to make the present scheme applicable to higher dimensions. 

\section{Scheme of the algorithms}
The scheme to search for various possible MEPs in general systems is composed of two steps: First, generate initial points ${\bm x}_{0}$ that are connected to some saddle points through the valley lines of the potential surface. Second, evolve the walkers from the initial distribution $q({\bm x}, 0)=\delta({\bm x}-{\bm x}_{0})$ according to the abovementioned procedure. In the previous two dimensional case~\cite{Akashi}, the first step was apparently easy, but in general dimensions, where the many-dimensional potential landscape cannot be straightforwardly visualized, we need a special treatment to specify appropriate ${\bm x}_{0}$. We therefore develop a method for this task, according to a principle that is obviously applicable to many dimensions. Afterwards, a simple and efficient parallel implementation of the second time evolution step is introduced. The success rate of tracking of the MEPs, defined with the frequency to generate the trajectories to the saddle point without straying from the proper paths, has been shown to depend strongly on the parameter $\delta$ in the biasing potential as $V({\bm x})=(1-\delta)U({\bm x})$, as well as the number of walkers $N_{\rm w}$. We also explore efficient control of $\delta$ to gain high success rate with relatively small $N_{\rm w}$.

\subsection{Generation of initial atomic positions}
In many dimensional cases the ``valley lines" of the potential can be formally defined as the lines generated by tracking upward the directions of local minimal potential gradient. That is why the walkers on a valley line are expected to relax more slowly to the minimum than those off the lines. This principle is demonstrated by a molecular mechanics simulation [Eq.~(\ref{eq:SDE-x})] for the 2D case with random initial positions and low temperature (see Supplementary movie). During the relaxation, the walkers drift to the stable energy minima in two stages. At first, the walkers run to the valley lines and then drift along the lines to the minimum slowly. Of course, the walker velocities depends on potential surface, but the relative velocities are slower on the valley lines. We assume that this simple principle applies to many-dimensional cases: The slowest walkers in the relaxation from random distribution must be on the reaction path.

A simple algorithm to generate an initial position ${\bm x}_{0}$ according to the above principle is to execute usual molecular mechanics simulation with relatively high temperature $T_{\rm high}$ for a while, set a threshold energy $E_{\rm th}$, relax the walkers with the heat bath quenched and specify the walker which comes down to $E_{\rm th}$ last. With this algorithm, however, it is obvious that only one valley line related to the smallest descent valley line can be tracked. To gain the diversity of the MEPs generated in the later time evolution step, we need to consider the different speeds of relaxations along various paths. For this purpose, we define atomic energy $E_{a} (a=1, 2, \dots, N_{\rm atom})$ with the interatomic potential $U_{ab}$ by $E_{a}=\sum_{b}U_{ab}$ and characterize the valley lines by the atoms showing largest energy deviations $\Delta E_{a}\equiv E_{a} - E_{a}^{\rm min}$ ($E_{a}^{\rm min}$: value of $E_{a}$ with the optimized atomic configuration). The background consideration of this strategy is as follows: Any reaction starts from the rupture of interatomic bonds. The potential energies between neighboring atoms are increased during the breaking of the bond. The reaction paths are therefore characterized by the combination of atoms showing largest $\Delta E_{a}$.
We have thus reached to the general algorithm to generate the initial positions that are related to diverse reaction paths, as summarized below (Algorithm 1). The algorithm has two stages: the first is to heat the system and the second one is to cool it. Simply speaking, we group the walkers according to the atoms showing $n$th largest $\Delta E_{a}$ and find the most slowly relaxing walkers among those groups. 

\begin{algorithm}[H]
\caption{Generation of the initial atomic positions as entrances to MEPs}
\begin{algorithmic}[1]
\scriptsize{
\STATE Set $T=T_{high}$   
\STATE Set $V({\bf x})=0$     ! Without biasing potential
\STATE Set $E_{cut}$ ! To set cut off energy $E_{cut} >  E_{min}$, $E_{min}$ is a minimum energy of a stable structure
\STATE Make the Langevin dynamics simulation for each walker $iw$ independently    ! To heat the walkers
\STATE Calculate the energy of each walker $E_{iw}$
\STATE To delete the walker $jw$ with $E_{jw} <  E_{cut}$  ! To make the gap between current energies and $E_{min}$
\STATE $N_{walker} = N_{walker} - N_{Deleted\_walkers}$  ! Recalculate the number of walkers
\STATE Set $T=T_{Low} \approx 0$   ! To cool the remained walkers to the stable structure with $E_{min}$
\STATE Set $E_{threshold} \approx E_{min}$   ! To catch the walker close to the stable state
\STATE ! The cooling Langevin dynamics
\STATE
\algdo  {$\text{   } t_i=0$, $\tau_{max}$ with $dt$}  ! Make the Langevin dynamics simulation for each walker $iw$ independently 
\STATE Save $\tau^{iw}_1 = t_i$ when $E_{iw} = E_{cut}$            ! The time of crossing $E_{cut}$
\STATE Save $\tau^{iw}_2 = t_i$ and atomic energies of walker with positions ${\bf x}_{iw}$ ($E^{iw}_{at}, at \in [1,N_{atoms}]$) when $E_{iw} = E_{threshold}$ 
\STATE Calculate $\Delta \tau^{iw} = \tau^{iw}_2 - \tau^{iw}_1$
\STATE 
\algend
}
\STATE Set $E^{min}_{at}$ ! Atomic energies at stable state, so $E_{min} = \sum_{at=1}^{N_{atoms}} {E^{min}_{at}}$
\STATE Calculate $\Delta E^{iw}_{at} = E^{iw}_{at} - E^{min}_{at}$ for each walker $iw$
\STATE Rank the energy deviations $\Delta E^{iw}_{at}$ for each walker $iw$ 
\STATE Calculate the numeric order of first $n$ atoms $\Theta^{iw}_{Order}$ with the highest energies (for each walker $iw$)
\STATE Make clusters with the same $\Theta^{iw}_{Order}$
\STATE Choose the slowest walker in each cluster $\Theta^{iw}_{Order}$ with a maximum $\Delta \tau^{iw}$  
\STATE Save the atomic coordinates ${\bf x_{iw}}$ the slowest walker in each cluster $\Theta^{iw}_{Order}$  ! There are initial atomic positions for reaction paths
\end{algorithmic}
\end{algorithm}

\subsection{Parallel implementation of the time evolution with strict conservation of $N_{\rm w}$}
As we explained above, the time evolution operator Eq.~(\ref{eq:timestep-decompose})
consists of the Langevin and the replication/removal steps. A simplistic implementation of the latter suffers from the fluctuating number of walkers $N_{\rm w}$, which is not suitable for efficient parallelization of the code. We therefore adopt an exactly number-conserving copying/removal algorithm in Ref.~\cite{conversation_walkers}, where this algorithm is named "[eq]". Let us here append the algorithm
\begin{enumerate}
\item
For the respective walkers (indicated by index $iw$) calculate
\begin{eqnarray}
\tilde{p}_{iw}
=
\frac{p_{iw}}{\sum_{jw} p_{jw}}
;\ \ 
p_{iw}
=
{\rm exp}\{L_{\rm rate}({\bf x}^{(iw)}, t) \tau \}
.
\label{eq:accum-prob}
\end{eqnarray}
$p_{N_{w}+1}$ is defined as $1$ for convenience.

\item
Generate a random number $d \in [0, 1/N_{w})$. 

\item
For each walker $iw$, increment $jw$ to find the range so that $\alpha_{iw}\equiv d+(iw-1)/N \in [\tilde{p}_{jw}, \tilde{p}_{jw+1})$; the $iw$th walker then inherits the position ${\bf x}^{(jw)}$.

\end{enumerate}
Through this algorithm, in the $N_{w}\rightarrow \infty$ limit the whole distribution function $q({\bf x},t)$ is modified to $q({\bf x},t+\tau) \propto {\rm exp}\{L_{\rm rate}({\bf x}^{(iw)},t) \tau \} q({\bf x}, t)$ with its norm $\int d{\bf x} q({\bf x},t)$ kept constant.

We have implemented a flat-MPI code for this algorithm. The pseudocode is as follows:
\begin{algorithm}
\caption{Parallelized number-conserving copying/removal}
\begin{algorithmic}[1]
\scriptsize{
\STATE Distribute the $N_{w}$ walkers to $N_{proc}$ processors and $N_{walker}=N_{w}/N_{proc}$
\STATE n1 = myrank*$N_{walker}$ + 1  \ \ \     ! Smallest walker index for each process
\STATE n2 = n1 + $N_{walker}$ - 1    \ \ \         ! Largest walker index for each process
\STATE prob(1:$N_{walker}$) = exp(tau*L\_rate(1:$N_{walker}$))      ! calculate the factor in Eq.~(\ref{eq:accum-prob})
\STATE p\_accum\_proc(n1) = prob(1)
\STATE
\algdo \  {iw=2, $N_{walker}$}
\STATE p\_accum\_proc(n1+iw-1) = p\_accum\_proc(n1 + iw -2)+prob(iw)
\STATE
\algend
\STATE call MPI\_ALLREDUCE(pr\_accum\_proc, pr\_accum, MPI\_SUM) ! Construct a shared array $\tilde{p}$ in Eq.~(\ref{eq:accum-prob})
}
\STATE Make the array of the accumulated $\tilde{p}$ by normalizing pr\_accum
\STATE Generate random number $d \in [0, 1/N_{w})$
\STATE call MPI\_BCAST($d$) \  ! Share $d$ for all the processes
\STATE \algdo {$\text{   } iw=1, N_{walker}$}
\STATE calculate $\alpha_{iw}\equiv d+(iw-1)/N$
\STATE Search $jw$ which satisfies $\alpha_{iw}\in [\tilde{p}_{jw}, \tilde{p}_{jw+1})$
\STATE \algend
\end{algorithmic}
\end{algorithm}

\subsection{Parameter settings for successful tracking}
In the previous paper it has been found that, during the time evolution, the center of $q({\bm x}, t)$ has fluctuations and the path search sometimes failed: the walkers stray normal to the MEP and move up the potential slope to a local maximum. This failure is understandable as the case where a large fraction of walkers are accidentally driven normal to the valley due to the random effect of the thermal fluctuation and copying/removal step. To reduce the probability of such failure with feasible number of walkers $N_{\rm w}$, we here consider appropriate setting of the parameters governing the simulation; the time step divided by the friction constant $dt/\Gamma$, temperature $T$ involved in the thermal fluctuation as well as the rate term, and $\delta$ entering the biasing potential as $V({\bm x})=(1-\delta)U({\bm x})$. 

Temperature $T$ determines the spread of $q({\bm x}, t)$ and its small spread allows us to describe the whole distribution accurately with a small number of walkers. On the other hand, too small $T$ induces radical ${\bm x}$ dependence of the rate term, as it enters $L_{\rm rate}$ in Eq.(\ref{eq:timestep-decompose}) as $\beta$. This effect can be mitigated by setting $dt/\Gamma$ small enough. A simple principle should therefore applies to $T$ and $dt/\Gamma$: Smaller is better (though they must be compromised not to make the evolution too slow). 

The effect of $\delta$ is more complicated~\cite{Akashi}. The sign of the potential force in the modified Langevin step [Eq.~(\ref{eq:Lang_step})] is inverted from the original $U({\bm x})$ with $\delta < 1/2$, which is the condition needed for the center of $q({\bm x}, t)$ to climb up the slope in a short time. Also, the speed of increase of the center and spread are both proportional to $1/\sqrt{\delta}$, which implies that, with too small $\delta$, the fluctuation of the walker distribution from the ideal $q({\bm x}, t)$ could be amplified with time. According to our preliminary analysis, the probability of failure is actually increased with a time evolution and tends to be larger with smaller $\delta$.

In order to avoid the fluctuation problem, we set $\delta$ as near to $1/2$ as possible and introduce a ``reset and pullback" step by some frequency; namely, ${\bm x}$ is set to their sample average for all the walkers to reset the sample distribution to the delta function, and later we execute a short simulation without $V({\bm x})$ to make sure that the distribution is well on the valley line. 

The whole procedure to track the MEPs from a given initial position, with the reset and pullback steps, is summarized as Algorithm~3. The performance of the current treatment is examined later for the Lennard-Jones cluster with seven particle (Sec.~\ref{sec:success}).
\begin{algorithm}
\caption{Simulation procedure with the reset and pullback steps}
\begin{algorithmic}[1]
\scriptsize{
\STATE Initialization
\STATE $N_{restart}$ is a number of the reset-pullback steps
\STATE $\delta = \delta_{initial} - 0.02$ ! Change delta smaller gradually to make $V({\bf x})$ stronger
\STATE
\algdo \  {N=1,$N_{restart}$}
\STATE Positions of all walkers equal average positions   ! The reset
\STATE a few time steps come back $V({\bf x})=0$
\STATE  $\delta = 0.5 - (0.5 - \delta_{initial}) \cdot (N_{restart} - N + 1 ) / N_{restart}$
\STATE $V({\bf x})=(1-\delta) \cdot U({\bf x})$
\STATE \algdo \  {t=0, $t_{max}/N_{restart}$ with $dt$}
\STATE time evolution with Eq.~\ref{eq:timestep-decompose} 
\STATE If ($U({\bf x}) < U({\bf x_0})$) and ($t > 400 \cdot dt$) THEN GO TO STEP 3   ! The potential energy has to increase
\STATE Save the trajectory - average positions and it's atomic energies
\STATE
\algend \ {    ! End of $t$ cycle }
\STATE
\algend \ {    ! End of $N_{restart}$ cycle } 
\STATE ! The relaxation from saddle point for each walker independently
\STATE $V({\bf x})=0$  ! The simulation without biasing potential 
\STATE 
\algdo \  {t=0, $t_{max}$ with $dt$}
\STATE Time evolution with Eq.~(\ref{eq:SDE-x})
\STATE Save the trajectories of $N_{w}$ walkers ! Each walker goes independently to it's local or global minimum energy state
\STATE
\algend \ {    ! End of $t$ cycle } for relaxation procedure
}
\end{algorithmic}
\end{algorithm}



\section{Benchmark test for argon cluster}
Let us show the performance of the present method with the Lennard-Jones (LJ) clusters with 7-38 atoms. The LJ clusters are well-studied systems and a good target to check our method. We used the model parameters in Ref.~\cite{LJ1990}, which corresponds to the interatomic potential of argon atoms. In the present study, we did not introduce the cutoff length and therefore numerical cost to calculate all the component of the force vector and Laplacian of the potential for each walker is proportional to $N_{\rm atom}^2$. 

The numbers of the degrees of freedom are then 114 at maximum, and the assembly of the walker represents the distribution function $q({\bm x}, t)$ in the 114-dimensional space. We visualize the structures of the clusters with the sample average of the position $\langle {\bm x} \rangle_{q({\bm x}, t)}$.

\begin{figure}[H]
 \begin{center}
    \includegraphics[scale=0.3]{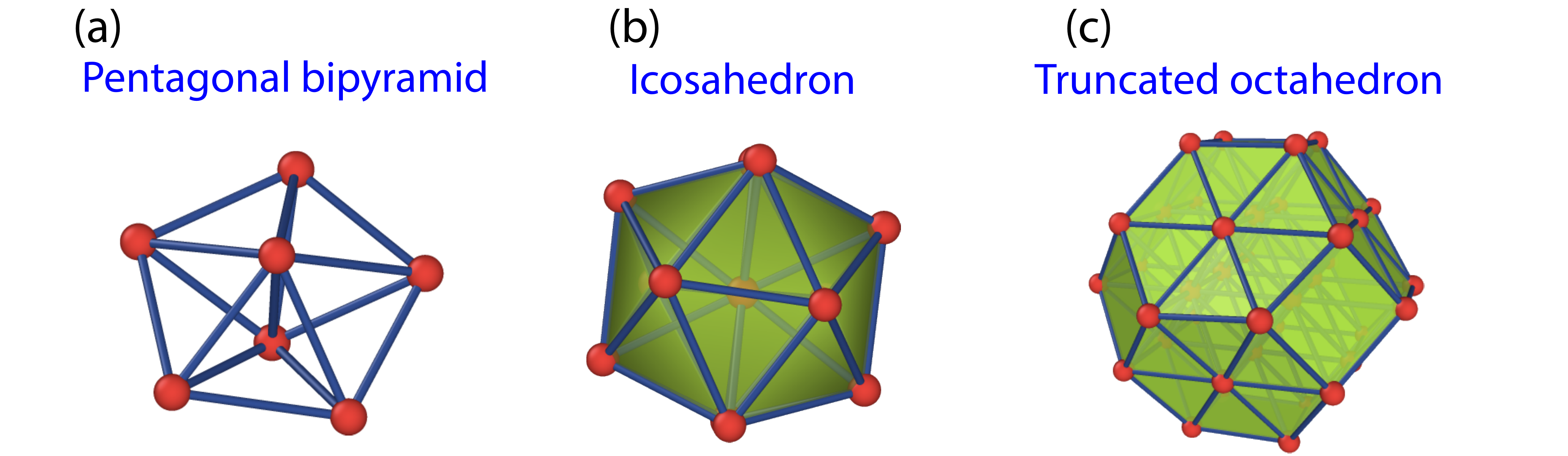}
    \caption{(Color online) Initial structures corresponding to energy minima: (a) LJ$_7$, (b) LJ$_{13}$, and (c) LJ$_{38}$.}
  \label{fig:LJ-initials}
 \end{center}
\end{figure}

In application of our scheme, we assume the situation where any local minimum structure (reactant) is known but the neighboring minima (products) are unknown. Here we start the simulations with the known minima as summarized in Fig.~\ref{fig:LJ-initials}, pentagonal bipyramid for LJ$_7$, icosahedron for LJ$_{13}$, and truncated octahedron for LJ$_{38}$, and seek for the possible MEPs connected from those. 

The simulation temperature in the MEP tracking process was $T=0.0001$ for LJ$_7$, $0.001$ for LJ$_13$ and LJ$_38$, respectively. The time step $dt/\Gamma=0.0004$ for all the simulations. The parameter $\delta$ was adaptively controlled as mentioned in Algorithm 3. The total number of walkers were $N_{w}=3200$ for the simulations in Secs.~\ref{sec:cluster7} and \ref{sec:cluster13-38}

\subsection{Parallel efficiency}

Here we summarize the scaling performance of our MPI-parallelized code. The benchmark calculation has been performed for the Lennard-Jones clusters with 7 and 38 particles (LJ$_7$ and LJ$_{38}$). The total number of walkers were set from 3456 to 345600, though tracking of the MEPs can be successful with far fewer walkers as discussed below. The calculations were performed with K computer, which contains 82944 compute nodes in total and each nodes has two 4-core SPARC64 VIIIfx processors.

\begin{figure}[H]
 \begin{center}
    \includegraphics[scale=0.85]{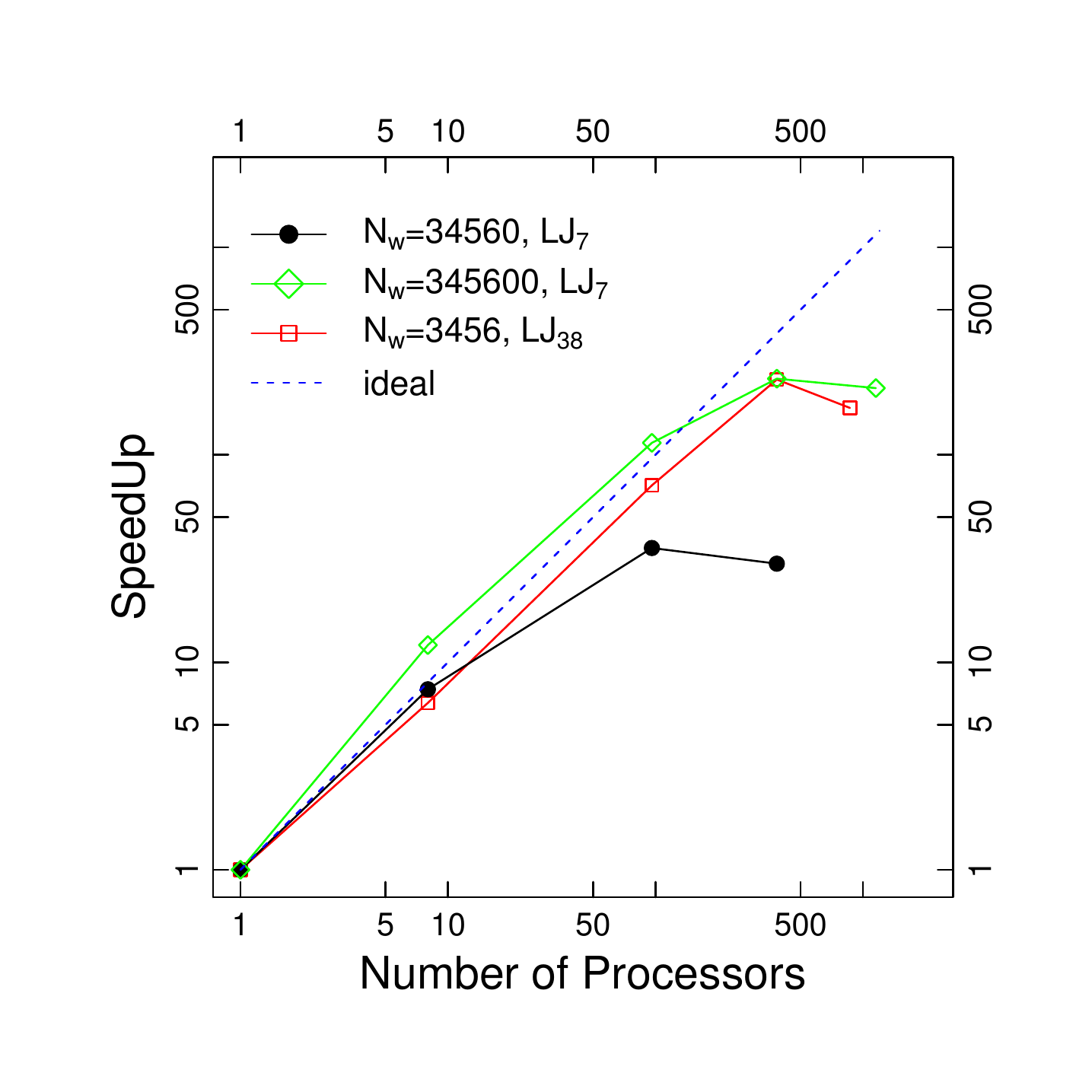}
    \caption{(Color online) Dependence of the speedup on the number of processes. Calculations were done with the number of processes 1, 8, 96, 384, 1152, so that the walkers are equally distributed to the processes.}
  \label{fig:speedup}
 \end{center}
\end{figure}

We measured the computational time for 100 timesteps and show the resulting speedup for various numbers of MPI processes in Fig.~\ref{fig:speedup}. There is a small excess of the speedup rate over ideal line in the region $N_{processors}<100$ for $N_w=345600$. It could be due to uniform memory distribution in the hard disk space for 1 processor with a huge number of walkers. 

In the case of LJ$_7$ the speedup saturated with the number of walkers per process around 300.
This relatively quick saturation is probably because of the moderate computational cost to calculate the potential and its derivatives in comparison with the MPI data exchange between processors. In fact the saturation of speedup rate for LJ$_{38}$ occurs when the number of walkers per process equals 36 and the computational cost to calculate the potential is increased by $(37 \cdot 38) / (6 \cdot 7) \approx 33.5$. Still, practically this performance was enough in this work for completing the single path tracking utilizing only 16 processors with 32000 time steps in 5 minutes for LJ$_{7}$ and in no more than several ten minutes for LJ$_{13}$. For LJ$_{38}$ we used 80 nodes (640 processes) with 80000 time steps in 7 hours. 

\subsection{Success rate}
\label{sec:success}
\begin{figure}[H]
\begin{minipage}[h]{0.5\linewidth}
\center{\includegraphics[width=1.0\linewidth]{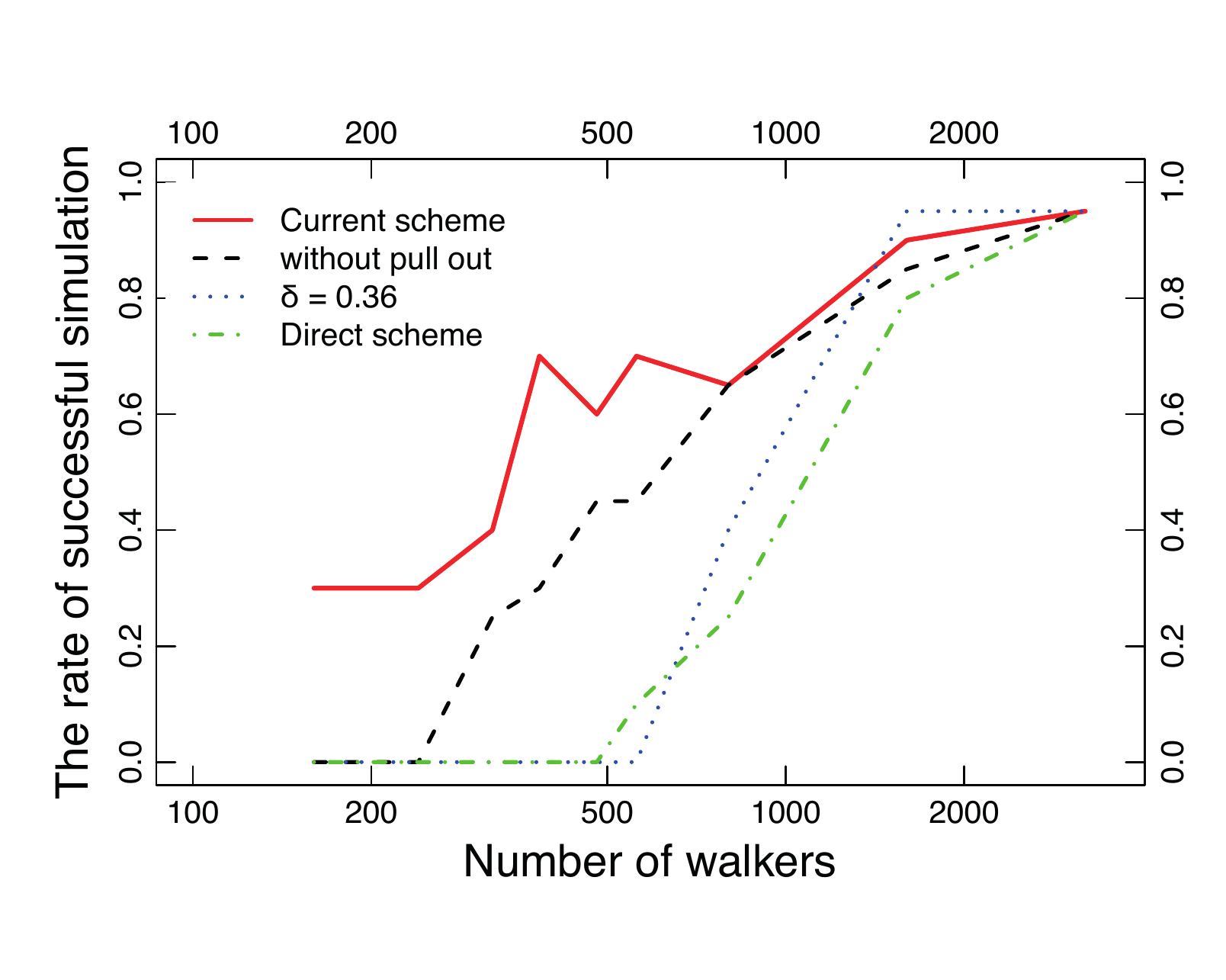}}
\end{minipage}
\hfill
\begin{minipage}[h]{0.5\linewidth}
\center{\includegraphics[width=1.0\linewidth]{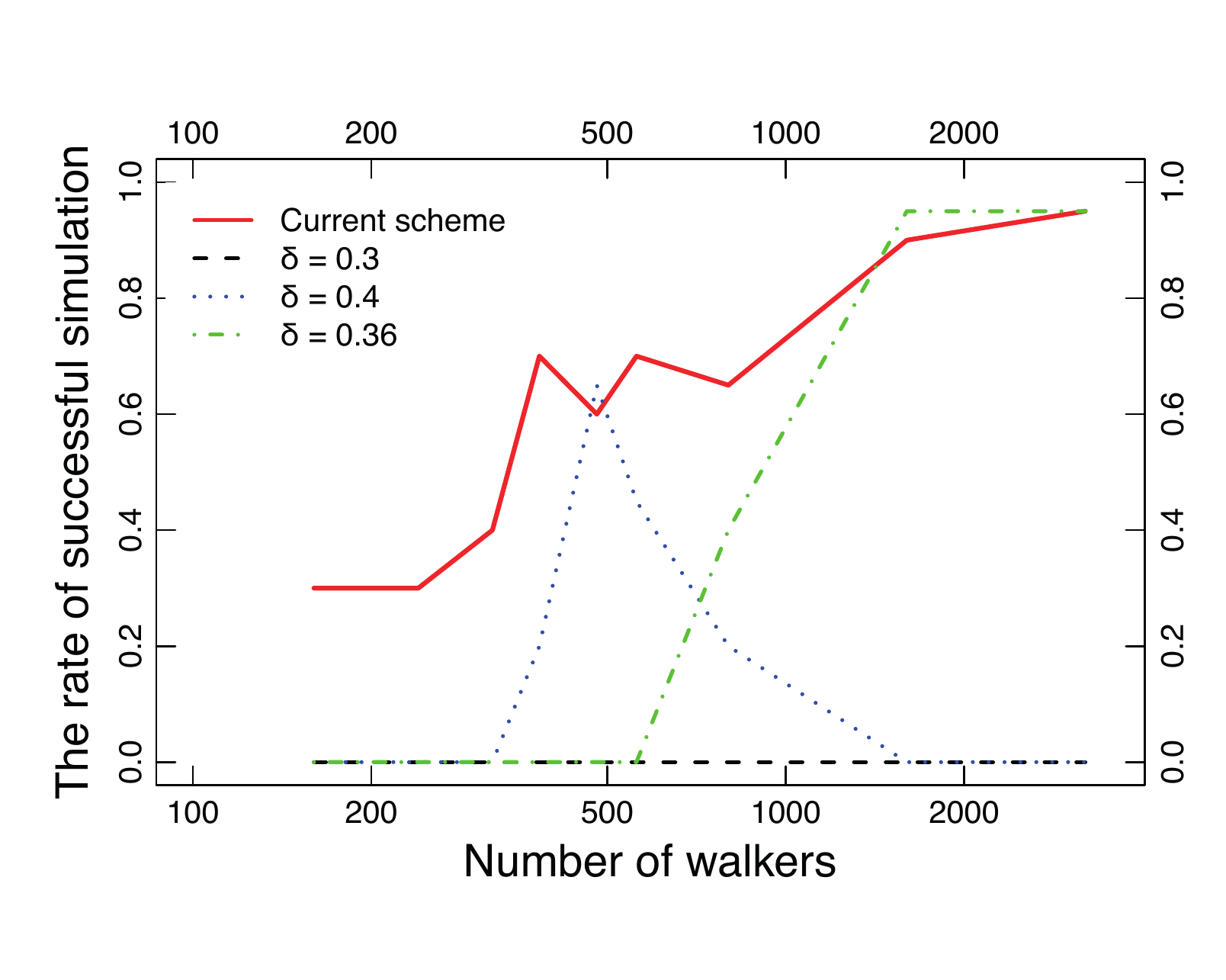}}
\end{minipage}
\hfill
\begin{minipage}[h]{0.5\linewidth}
\center{\includegraphics[width=1.0\linewidth]{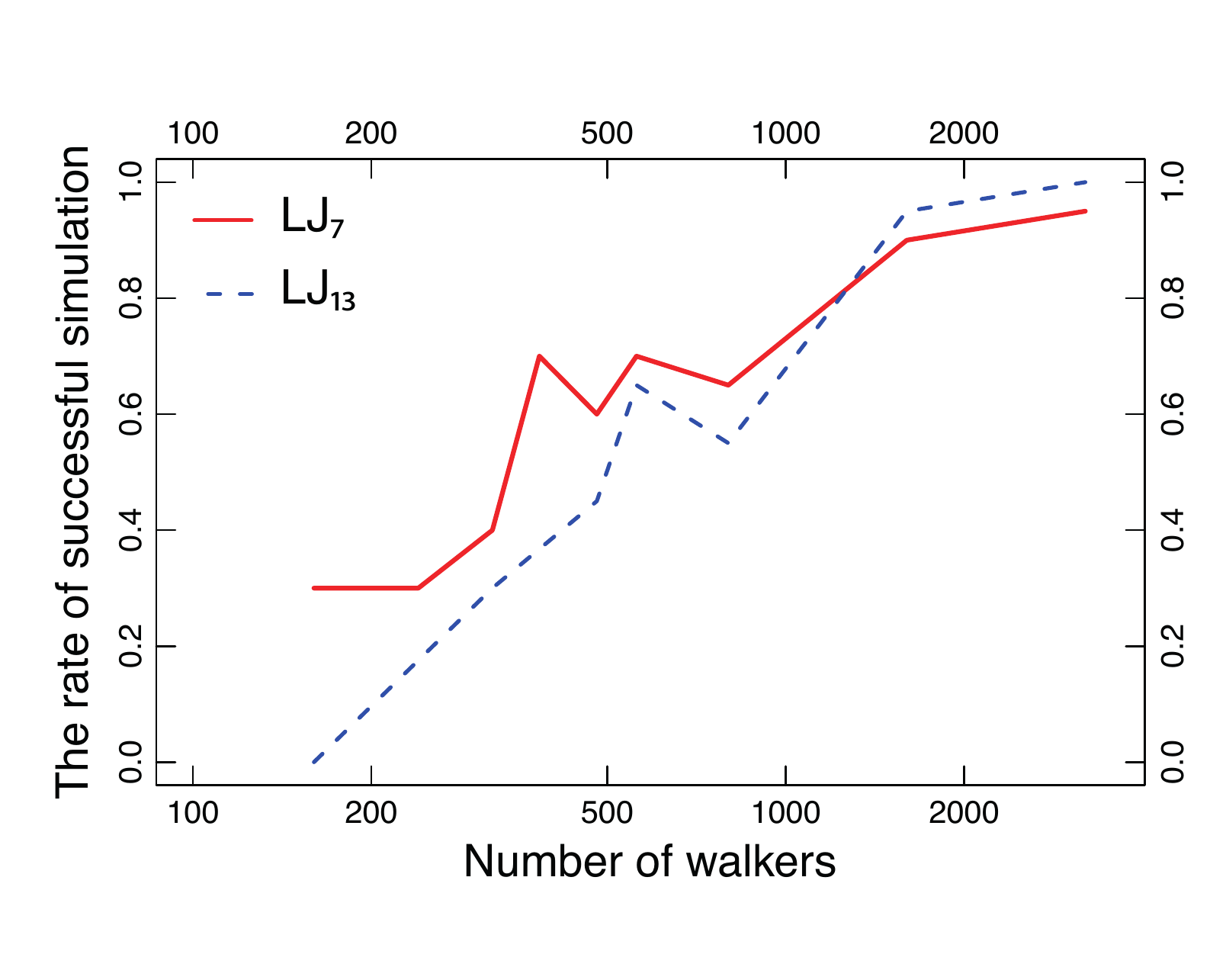}}
\end{minipage}
\caption{(Color online) The successful rate of simulations: the top left figure is measures for different scheme of simulations, the top right one is for fixed delta parameter, and bottom figure is for LJ$_7$ and LJ$_{13}$.}
\label{Succ_rate}
\end{figure}

To estimate the influence of $\delta$ controlling the strength of the biasing potential $V({\bm x})$ on a successful rate of simulation we repeated the simulation starting from a single initial position for LJ$_7$ and LJ$_{13}$ (Fig.~\ref{Succ_rate}) to count the number of successful tracking to the saddle point connected through the valley line. The results are compared with and without the reset and pullback steps and with various ways to control $\delta$. The number of walkers was 16, 160, 240, 320, 384, 480, 560, 800, 1600, 3200 and the number of repeating simulations was 20 for each point. For $N_w>3200$ all simulations were successful. 
It is clear that the most significant parameter in seeking the reaction paths is the delta parameter $\delta$. In the current scheme (Algorithm 3), the $\delta$ adaptive selection is performed by gradually increasing the influence of the biasing potential $V({\bf x})=(1-\delta) \cdot U({\bf x})$

On Fig.~\ref{Succ_rate} there is no successful simulation with $\delta=0.3$ because under strong V({\bf x}) all simulations go to the higher order saddle points and the system is destroyed into several parts. If the delta parameter is too near to $1/2$ like 0.4, then the V({\bf x}) is too weak and the walkers do not go far up the potential slope. It is interesting that for the small number of walkers $N_w < 1000$ the system is able to go out of balance due to fluctuations, an amplitude of which becomes higher with the small number of walkers. As a result, the system sometimes drifts up to the saddle point successfully. Of course, when the walker number is too small $N_w < 400$, the fluctuations become too large and the system is destroyed again.  

We also tried to calculate the reaction paths without the additional procedures; reset, pullback and adaptive $\delta$ (top left Fig.~\ref{Succ_rate}). For LJ$_7$, $\delta = 0.36$ was the minimal value for drifting up on the potential valley, that's why it has a good rate of successful simulations. The procedures of the reset and pullback are also important only for region $N_w < 1000$. We need to keep in mind that for successful simulation it is necessary to choose other parameters like time step $dt$, temperature $T$, simulation time $t_{max}$ and etc. From this point of view, short simulations in order to determine the parameter settings are generally essential. Remarkably, the number of walkers required for successful tracking of MEPs does not change so significantly with increasing the system dimension (bottom Fig.~\ref{Succ_rate}), For LJ$_7$, LJ$_{13}$ and LJ$_{38}$, $N_w=3200$--$6400$ was enough.

\subsection{Simulation results for argon cluster with 7 atoms}
\label{sec:cluster7}
\begin{figure}[H]
 \begin{center}
 \includegraphics[scale=0.75]{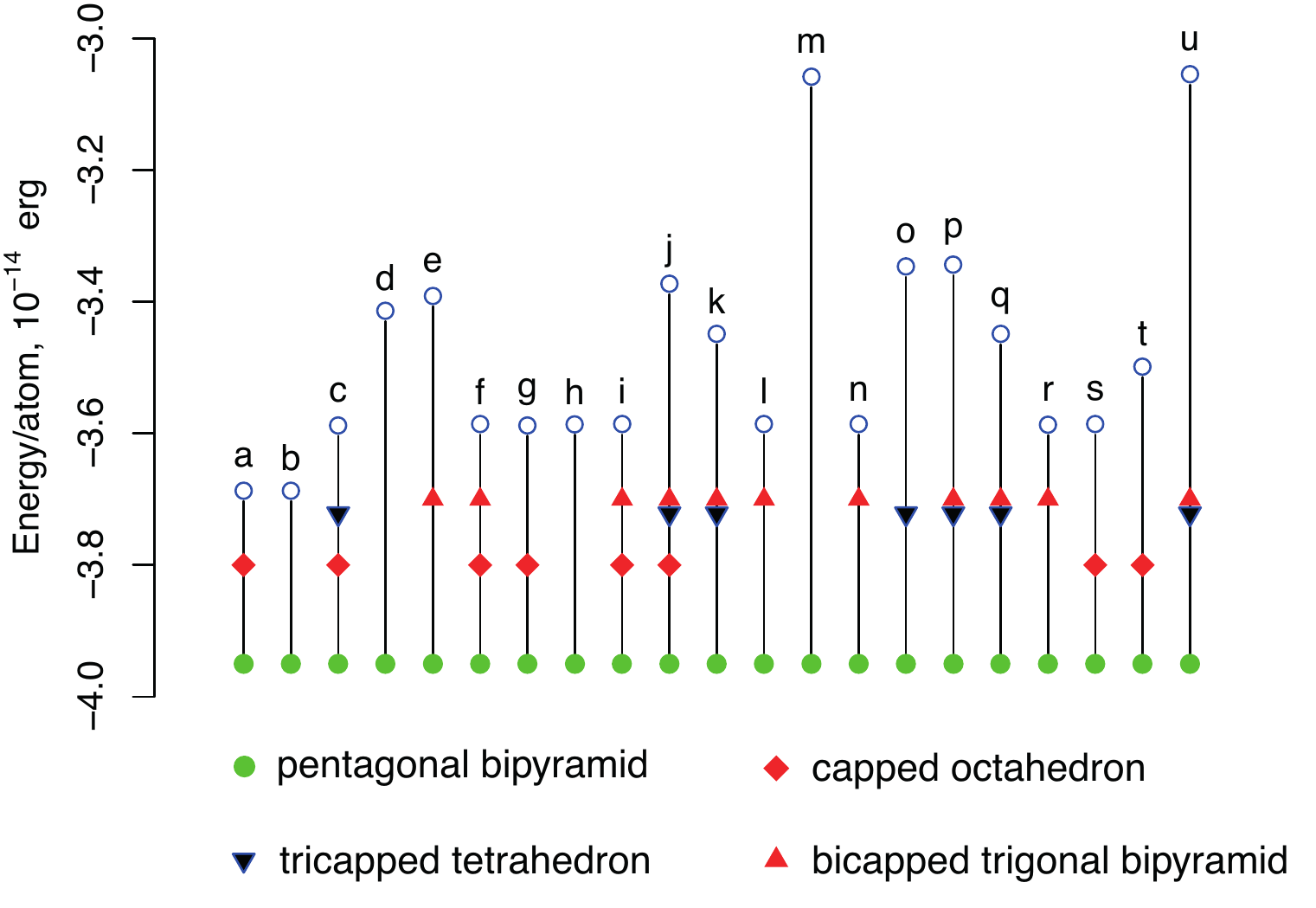}
  \caption{(Color online) Schematic view of the local minimuma and transition states for $Ar_7$ cluster potential energy surface.}
  \label{saddle_points7}
 \end{center}
\end{figure}

Utilizing the algorithm of seeking the initial atomic positions (Algorithm~1) for LJ$_7$, we have determined 42 initial positions ${\bm x}_{0}$ as entrances to the MEPs from one stable state - pentagonal bipyramid. It should be noted only the simulations from 2 initial positions did not reach the saddle point (SP) in 32000 time steps, but the other 40 successfully yielded the paths to the SP. 
The LJ$_7$ has $C_{5h}$ group symmetry as shown in Fig.~\ref{fig:LJ-initials}, that's why some simulations have the same trajectories. 
As a result, there are only 21 unique pathways (Fig.~\ref{saddle_points7}). We define the unique paths as a way through the unique SP and unique SP by the labels of atoms showing 1st to 3rd largest energy deviations $\Delta E_{a}$ with the corresponding ${\bm x}_{0}$ (atomic energy spectrum of SP).  

\begin{figure}[H]
\begin{minipage}[h]{1.0\linewidth}
 \begin{center}
 \includegraphics[scale=0.7]{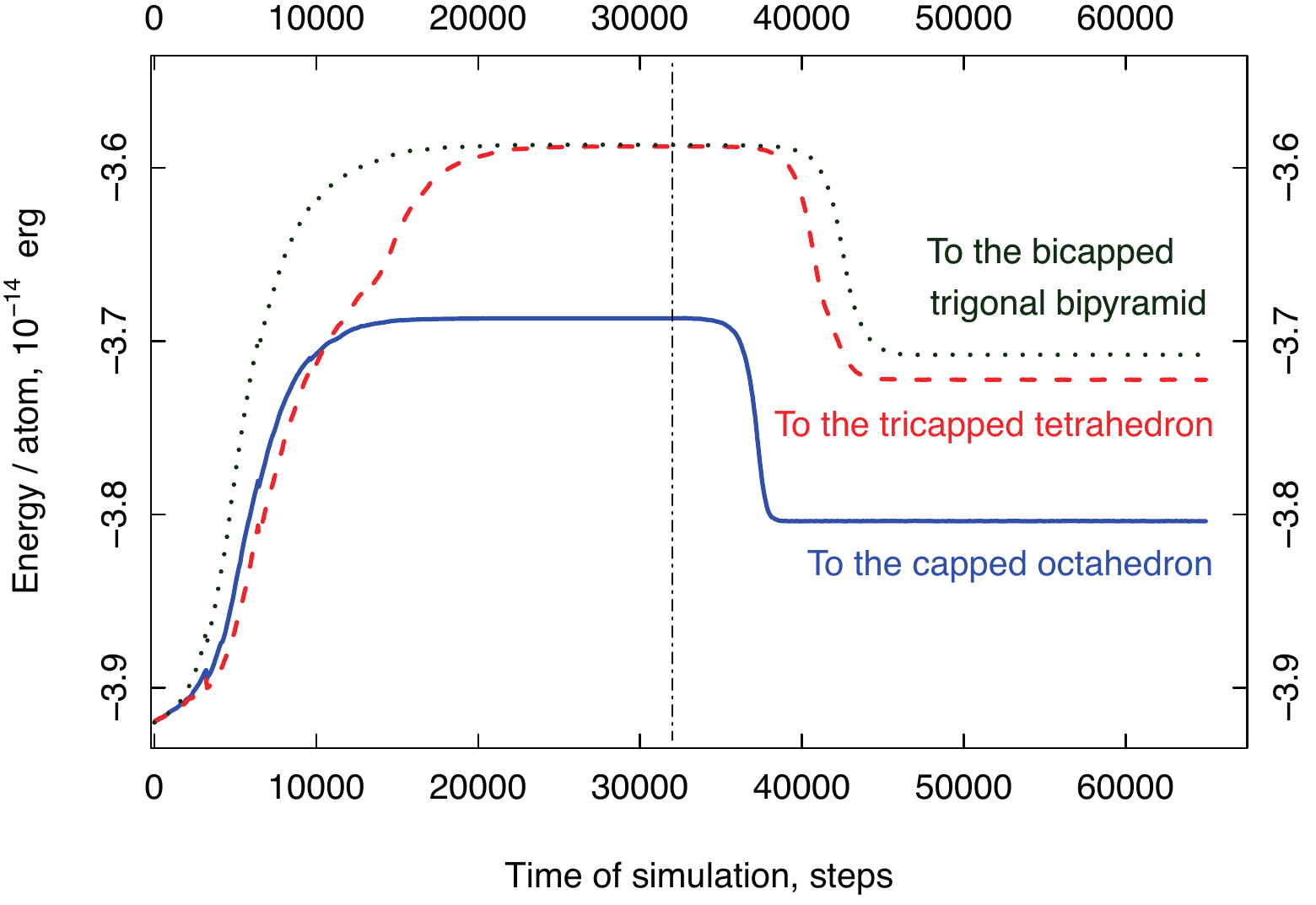}
  \caption{(Color online) The energy evolution of reaction paths for a argon cluster LJ$_7$ during simulation under biasing potential and during the relaxation. The horizontal line shows the moment when the relaxation start off.
    \vspace{\baselineskip} }
  \label{energy7}
   \end{center}
 \end{minipage}
\begin{minipage}[h]{1.0\linewidth}
 \begin{center}
 \includegraphics[scale=0.235]{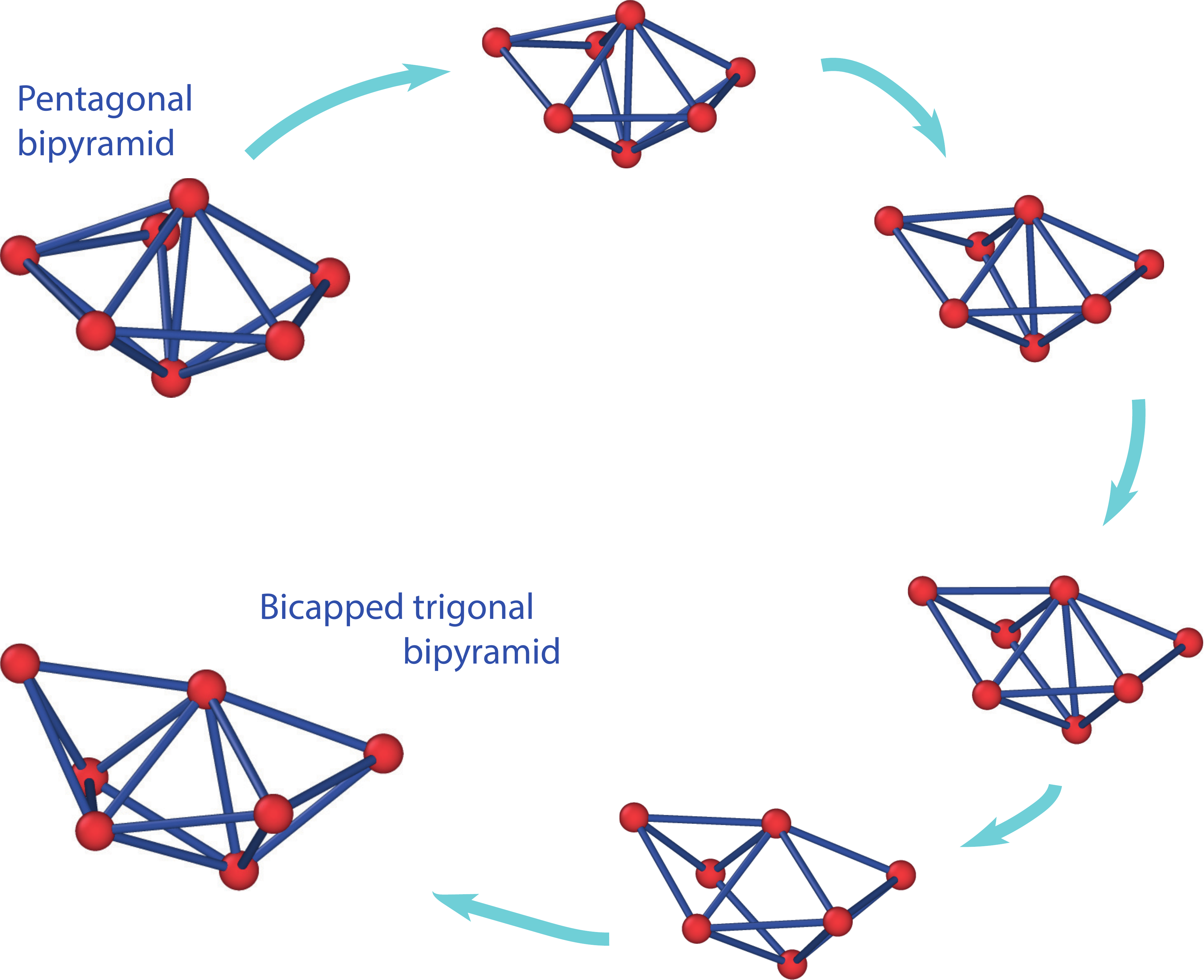}
  \caption{(Color online) The snapshots of reaction path from pentagonal bipyramid to the the bicapped trigonal bipyramid structure for the argon cluster LJ$_7$.}
  \label{3gonal}
 \end{center}
 \end{minipage}
\end{figure}

The unique 21 SPs are shown in (Fig.~\ref{saddle_points7}). The curves of the potential energy along the paths to other metastable points are shown in Fig.~\ref{energy7}, which goes through the a, c and l saddle points whose energy barriers agree with those determined in Ref.~\cite{LJ1990}. The corresponding reaction paths are visualized in Figs.~\ref{3gonal},~\ref{octahedron},~\ref{3capped}.
In the Supplementary movies, we have shown the dynamics of reaction paths during the simulations. 

There are a lot of studies of reactions in LJ clusters, for example \cite{LJ1999_38,LJ1999_size,LJ2000_38,LJ2006,LJ2018_38}. Almost all papers are concentrated on the topic of free-energy landscape and their energy minima, using the different methods: molecular dynamics \cite{LJ1999_38,LJ1999_size}, parallel tempering technique with molecular dynamics \cite{LJ2000_38}, path-sampling scheme \cite{LJ2006} and diffusion Mote-Carlo method \cite{LJ2018_38}. The free-energy landscape is defined in a coarse-grained low-dimensional space with collective order parameter axes like $Q_4$ or $Q_6$. In our method, on the other hand, we directly generated the MEPs keeping all the atomic degrees of freedom. We also note that, differently from the existing methods using non-empirical potential force like metadynamics~\cite{Laio01102002}, we use the original potential $U({\bm x})$ to keep the landscape unmodified.

\begin{figure}[H]
\begin{minipage}[h]{1.0\linewidth}
 \begin{center}
 \includegraphics[scale=0.25]{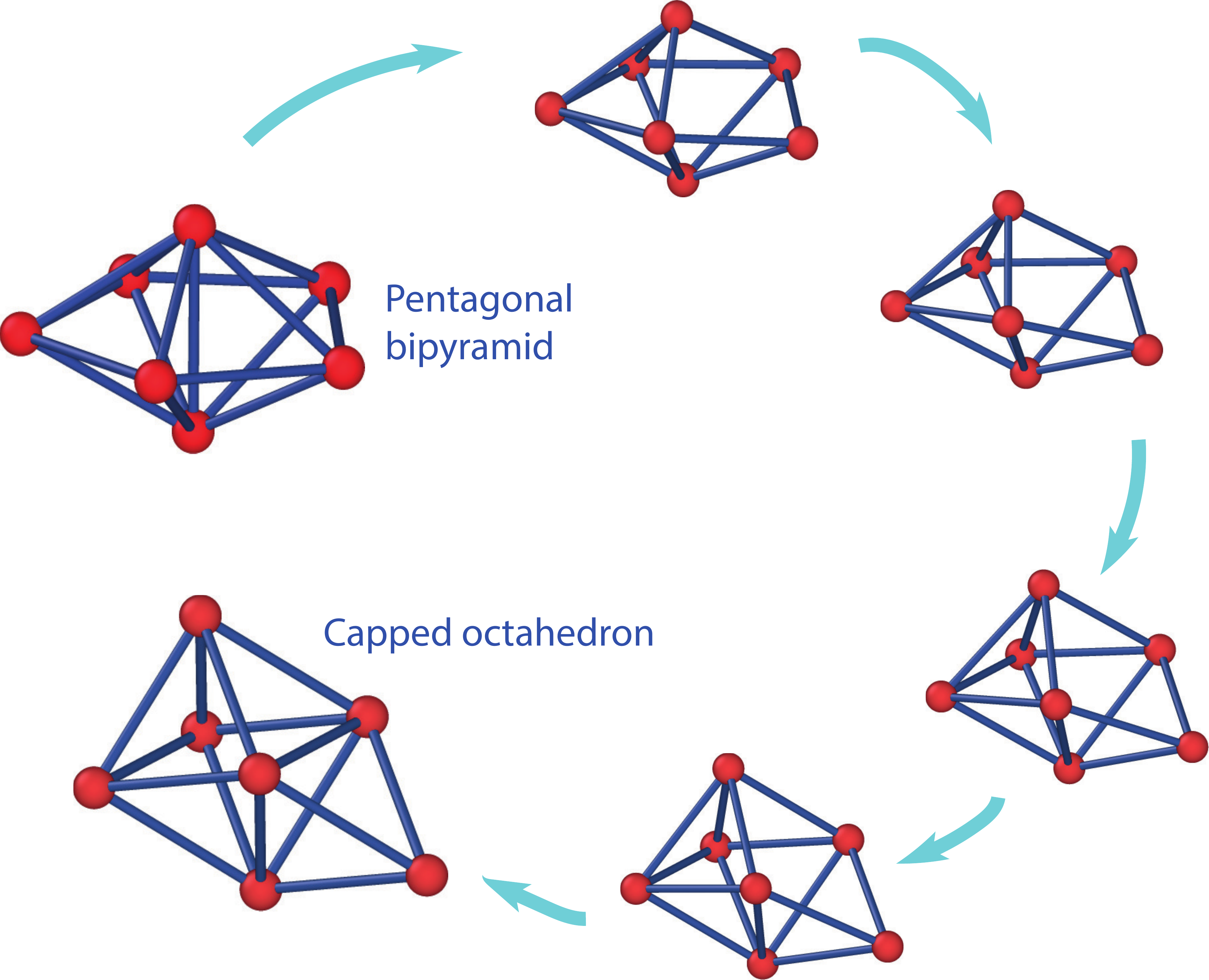}
  \caption{(Color online) The snapshots of reaction path from pentagonal bipyramid to the capped octahedron structure for the argon cluster LJ$_7$.
  \vspace{\baselineskip}  }
  \label{octahedron}
 \end{center}
\end{minipage}
\begin{minipage}[h]{1.0\linewidth}
 \begin{center}
 \includegraphics[scale=0.24]{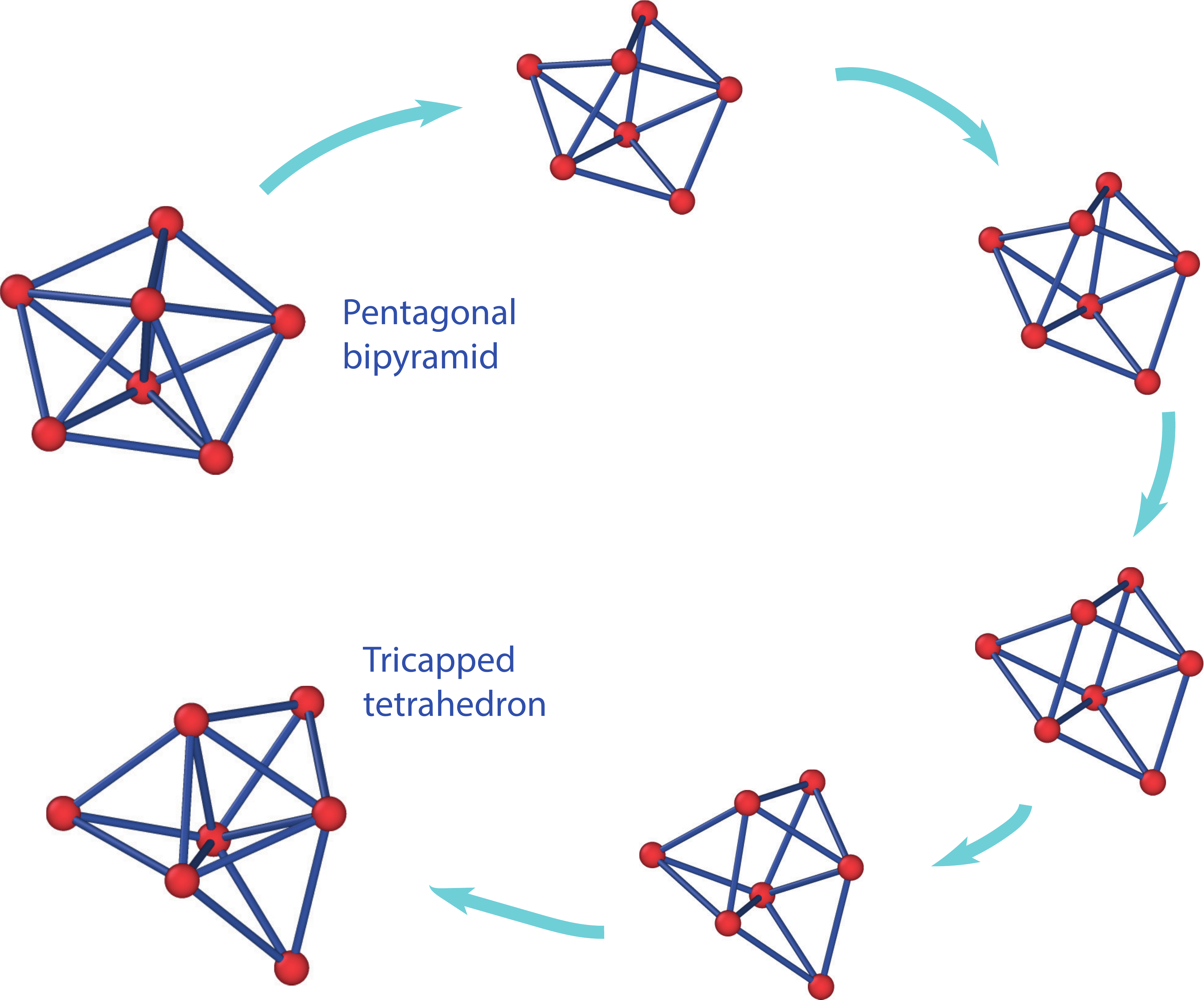}
  \caption{(Color online) The snapshots of reaction path from pentagonal bipyramid to the tricapped tetrahedron structure for the argon cluster LJ$_7$.}
  \label{3capped}
 \end{center}
 \end{minipage}
\end{figure}

An interesting advantage of our method is possibility to reach high energy barrier as easy as low energy one. 
In comparison with the previous study~\cite{LJ1990}, where authors found the 4 saddle points from the LJ$_7$ pentagonal bipyramid, our method have found the 21 SPs (Fig.~\ref{saddle_points7}). The 21 SPs include the first order SP (a,b,c,l,n,r points in Fig.~\ref{saddle_points7}) and high order SPs.

\subsection{Simulation results for argon LJ$_{13}$ and LJ$_{38}$ clusters}
\label{sec:cluster13-38}
To check how the method behaves with increasing of dimensionality, we performed several simulations for  LJ$_{13}$ and LJ$_{38}$ clusters. The parallel efficiency is increased with a number of atoms (Fig.~\ref{fig:speedup}), but the total number of walkers is still constant around $3200$--$6400$ (bottom of Fig.~\ref{Succ_rate}). It seems possible to apply the method to huge systems with several hundred or even several thousands of atoms without increasing the number of walkers so much, and that is our future task. In our opinion the reason for this ``good" behavior of method is the fact that only several atoms take part in the reaction while a movement of other atoms is deterministic, with which the volume spanned by $q({\bm x}, t)$ is kept feasibly small.

\begin{figure}[H]
 \begin{center}
 \includegraphics[scale=0.75]{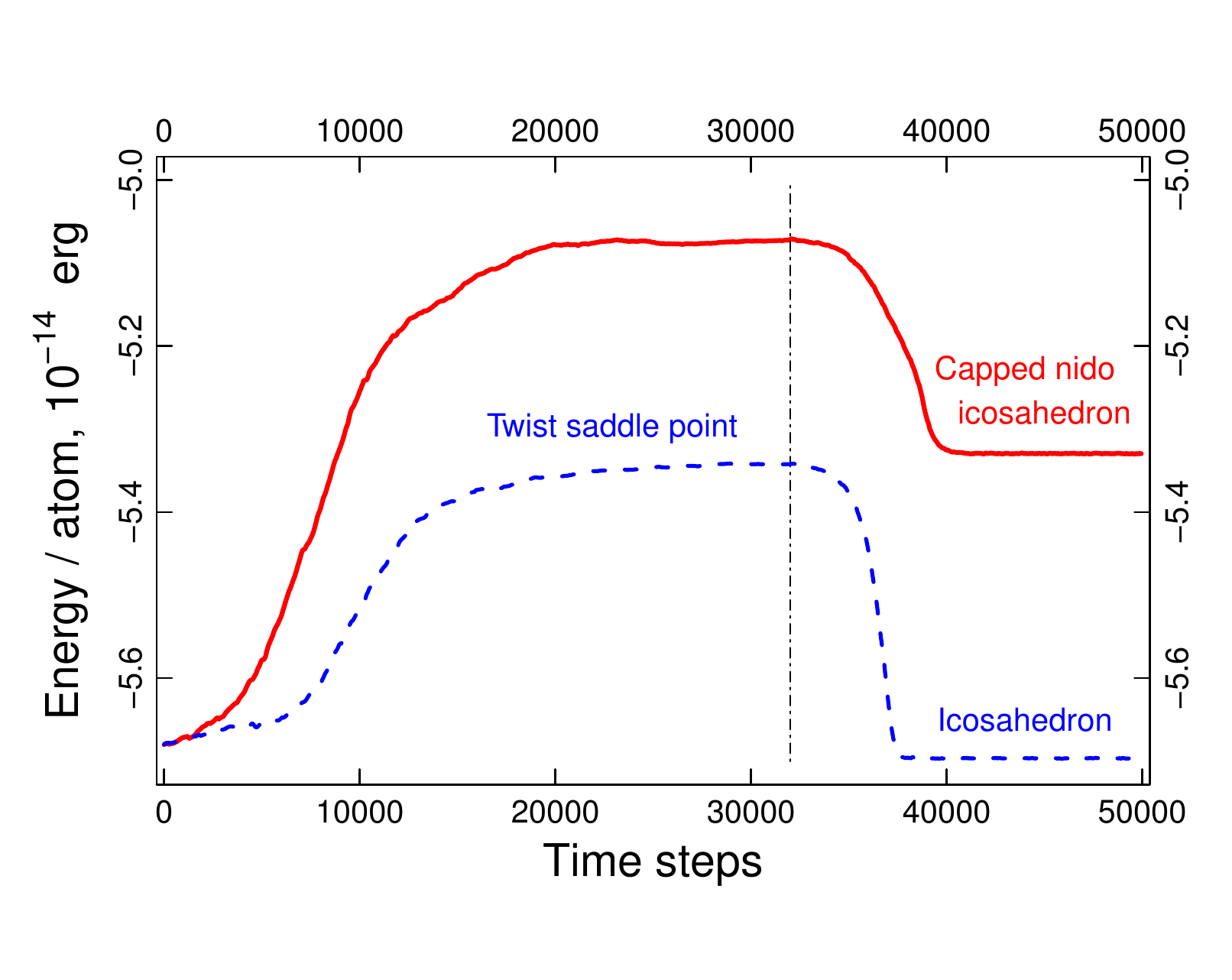}
  \caption{(Color online) The energy evolution of reaction paths for a argon cluster LJ$_{13}$ during simulation under biasing potential and during the relaxation. The horizontal line shows the moment when the relaxation start off.
  \vspace{\baselineskip}  }
  \label{13_energy}
 \end{center}
\end{figure}

\begin{figure}[H]
\begin{minipage}[h]{1.0\linewidth}
 \begin{center}
 \includegraphics[scale=0.25]{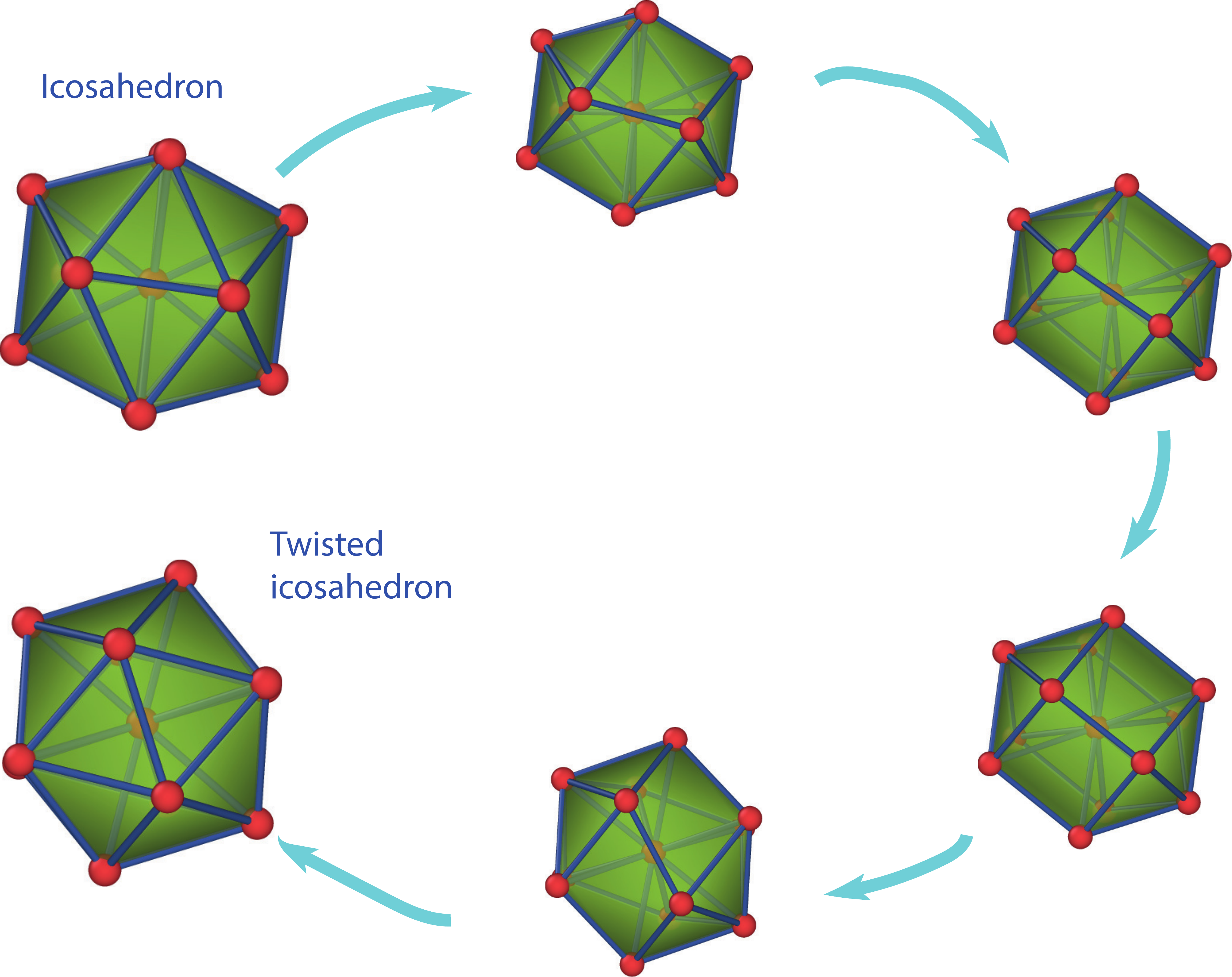}
  \caption{(Color online) The snapshots of reaction path from icosahedron structure to the same structure for the argon cluster LJ$_{13}$ due to the twisted process.
  \vspace{\baselineskip}  }
  \label{13_twist}
 \end{center}
\end{minipage}
\begin{minipage}[h]{1.0\linewidth}
 \begin{center}
 \includegraphics[scale=0.24]{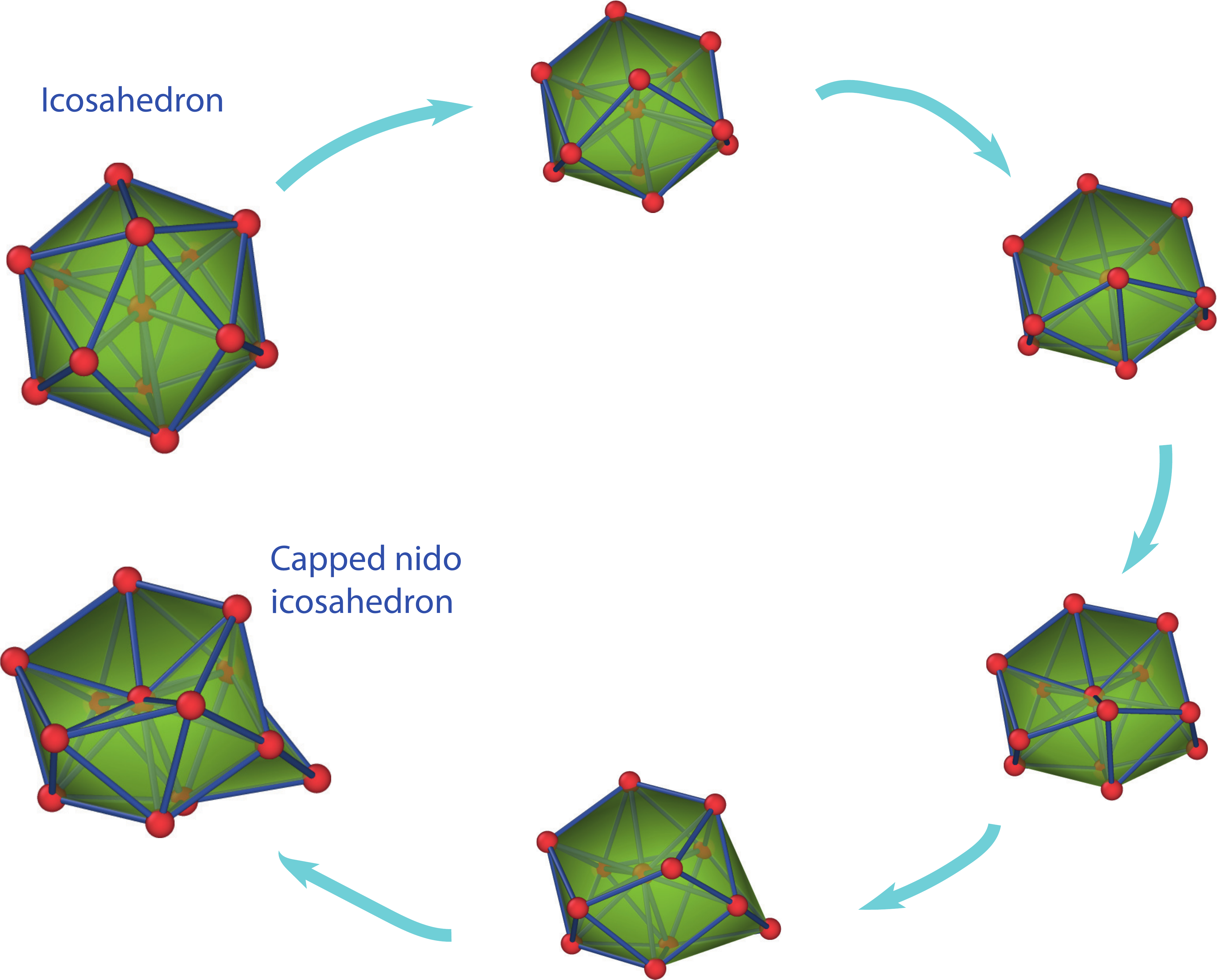}
  \caption{(Color online) The snapshots of reaction path from icosahedron structure to the capped nido icosahedron structure for LJ$_{13}$.}
  \label{13_path}
 \end{center}
 \end{minipage}
\end{figure}

\begin{figure}[H]
\begin{minipage}[h]{1.0\linewidth}
 \begin{center}
 \includegraphics[scale=0.75]{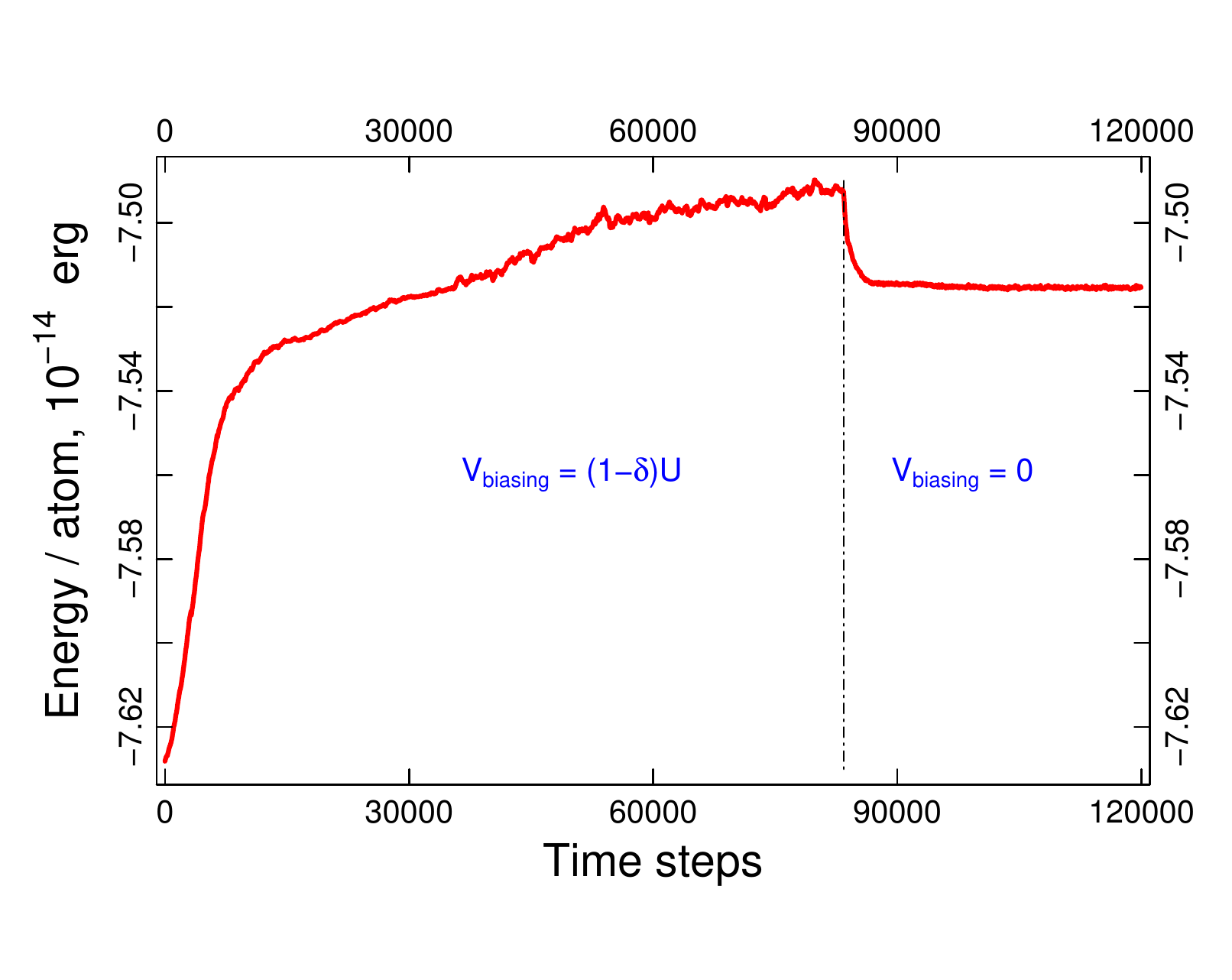}
  \caption{(Color online) The energy evolution of reaction path for LJ$_{38}$ during simulation under biasing potential and during the relaxation. The horizontal line shows the moment when the relaxation start off.
  \vspace{\baselineskip}  }
  \label{38_energy}
 \end{center}
\end{minipage}
\begin{minipage}[h]{1.0\linewidth}
 \begin{center}
 \includegraphics[scale=0.24]{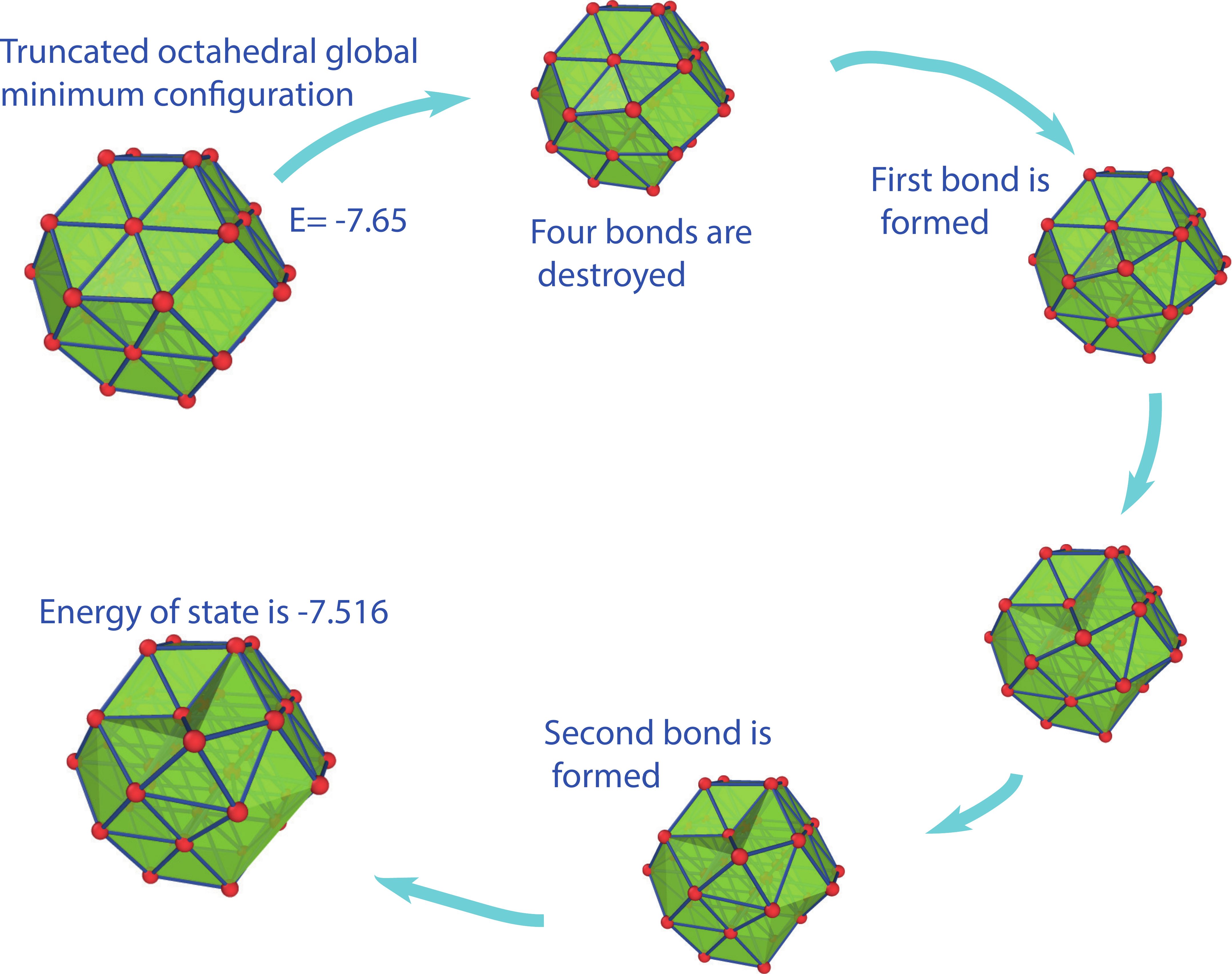}
  \caption{(Color online) The snapshots of reaction path for argon cluster LJ$_{38}$.}
  \label{38_path}
 \end{center}
 \end{minipage}
\end{figure}

Fig.~\ref{13_energy} shows the energy evolution of LJ$_{13}$ cluster for two reaction pathways from icosahedron: the first is to the twist SP, beyond which the same icosahedron is reached (Fig.~\ref{13_twist}), and the second is to the capped nido icosahedron (Fig.~\ref{13_path}). Notably, the previous study~\cite{LJ1990} didn't mention the twist SP, despite it is clear to see this structure as unstable and SP for icosahedron - icosahedron transformation with twisted movement (see Supplementary movie). 
The structure transformation is shown in Fig.~\ref{38_path}. The truncated octahedral configuration has very high symmetry and therefore corresponds to a deep energy global minimum, that is why we performed long simulation to reach to the SP: at first the 32000 time steps with a low temperature and, then we increased the $T$ in order to accelerate the simulation. 

For small clusters (like LJ$_{7}$) all reactions occur in accordance with diamond-square-diamond (DSD) process \cite{LJ1990}, where is usually one bond is destroyed and then another bond forms (Figs.~\ref{3gonal},\ref{octahedron},\ref{3capped}). For LJ$_{13}$ in the case of twisted reaction, there are 5 DSD processes at the same time (Fig.~\ref{13_twist}). Figure~\ref{13_path} shows the "icosahedron - capped nido icosahedron" transformation, which has a vacancy formation on the surface and DSD process induced by a capped atom. 

For  LJ$_{38}$ a reaction path is also complicated (see Supplementary movie): 
a pair of atoms simultaneously and synchronously change their coordinates during the reaction;
in this case, two DSD processes occur, and also the second atom breaks two bonds, forming a vacancy (Fig.~\ref{38_path}); 
the first atom forms a vertex, and the second is the base of a small pyramid on the surface of the initial configuration.
In addition, during the reaction path, we observed a collective displacement of all atoms towards the atoms forming the pyramid, which reached its maximum at the saddle point and disappeared in a stable state. The amplitude of atomic displacement depends on the distance from atom to reaction couple of atoms.
Using collective variable such detailed process is impossible to detect.

\section{Conclusion}

In this work we showed the concept, mathematical model and algorithmic steps of our method. The method is based on the Master equation (modified Smoluchowski equation), which is related to molecular mechanics described by the Langevin equation with weighted sampling. The model allows us to find the probability distribution through the biasing distribution, which is calculated under biasing potential. To solve the Master equation we use the stochastic walker algorithm with parallelized number-conserving copying/deletion. We verified the scaling performance of our MPI-parallelized code with a benchmark of LJ$_7$ and LJ$_{13}$ clusters. Using parallel calculation on K-computer we have shown that the speedup effectiveness increases with the dimension of the system. The computational cost is increased due to the calculation of the potential function and diagonal elements of Hessian matrix like $N_{atoms}^2 \cdot N_w$, but $N_w$ is constant for 7-38 atoms.

We have designed the algorithm to seek the initial positions as entrances to the MEPs and to track the MEPs efficiently (adaptive delta parameter, reset and pullback) procedures in order to keep the walkers on the MEPs. 
The successful performance of tracking the MEPs in the LJ cluster systems with various number of particles with the same number of walkers suggests that larger systems could be feasible with the present simulation scheme, paving a path to the simulation of reaction mechanics free from artificial potential or a priori collective coordinates.

\bigskip

{\bf Acknowledgment} {\itshape This research was supported by MEXT as Exploratory Challenge on Post-K computer (Frontiers of Basic Science: Challenging the Limits). This research used computational resources of the K computer provided by the RIKEN Advanced Institute for Computational Science through the HPCI System Research project (Project ID:hp160257, hp170244, hp180184).}


\newpage

\section{Refences}
\bibliographystyle{elsarticle-num}
\bibliography{CPC}

\begin{thebibliography}{10}
\expandafter\ifx\csname url\endcsname\relax
  \def\url#1{\texttt{#1}}\fi
\expandafter\ifx\csname urlprefix\endcsname\relax\def\urlprefix{URL }\fi
\expandafter\ifx\csname href\endcsname\relax
  \def\href#1#2{#2} \def\path#1{#1}\fi

\bibitem{Carter-Ciccotti-bluemoon}
E.~Carter, G.~Ciccotti, J.~T. Hynes, R.~Kapral,
  \href{http://www.sciencedirect.com/science/article/pii/S0009261489873142}{Constrained
  reaction coordinate dynamics for the simulation of rare events}, Chem. Phys.
  Lett. 156~(5) (1989) 472 -- 477.
\newblock \href
  {http://dx.doi.org/https://doi.org/10.1016/S0009-2614(89)87314-2}
  {\path{doi:https://doi.org/10.1016/S0009-2614(89)87314-2}}.
\newline\urlprefix\url{http://www.sciencedirect.com/science/article/pii/S0009261489873142}

\bibitem{Sprik-Ciccotti-bluemoon}
M.~Sprik, G.~Ciccotti, \href{https://doi.org/10.1063/1.477419}{Free energy from
  constrained molecular dynamics}, J. Chem. Phys. 109~(18) (1998) 7737--7744.
\newblock \href {http://arxiv.org/abs/https://doi.org/10.1063/1.477419}
  {\path{arXiv:https://doi.org/10.1063/1.477419}}, \href
  {http://dx.doi.org/10.1063/1.477419} {\path{doi:10.1063/1.477419}}.
\newline\urlprefix\url{https://doi.org/10.1063/1.477419}

\bibitem{Schlitter-Wollmer-targetedMD}
J.~Schlitter, M.~Engels, P.~Kruger, E.~Jacoby, A.~Wollmer,
  \href{https://doi.org/10.1080/08927029308022170}{Targeted molecular dynamics
  simulation of conformational change-application to the t -- r transition in
  insulin}, Mol. Simul. 10~(2-6) (1993) 291--308.
\newblock \href
  {http://arxiv.org/abs/https://doi.org/10.1080/08927029308022170}
  {\path{arXiv:https://doi.org/10.1080/08927029308022170}}, \href
  {http://dx.doi.org/10.1080/08927029308022170}
  {\path{doi:10.1080/08927029308022170}}.
\newline\urlprefix\url{https://doi.org/10.1080/08927029308022170}

\bibitem{SteeredMD-orig-Sci1996}
H.~Grubm{\"u}ller, B.~Heymann, P.~Tavan,
  \href{http://science.sciencemag.org/content/271/5251/997}{Ligand binding:
  Molecular mechanics calculation of the streptavidin-biotin rupture force},
  Science 271~(5251) (1996) 997--999.
\newblock \href
  {http://arxiv.org/abs/http://science.sciencemag.org/content/271/5251/997.full.pdf}
  {\path{arXiv:http://science.sciencemag.org/content/271/5251/997.full.pdf}},
  \href {http://dx.doi.org/10.1126/science.271.5251.997}
  {\path{doi:10.1126/science.271.5251.997}}.
\newline\urlprefix\url{http://science.sciencemag.org/content/271/5251/997}

\bibitem{Voter-hyperdyn-JCP1997}
A.~F. Voter, \href{https://doi.org/10.1063/1.473503}{A method for accelerating
  the molecular dynamics simulation of infrequent events}, J. Chem. Phys.
  106~(11) (1997) 4665--4677.
\newblock \href {http://arxiv.org/abs/https://doi.org/10.1063/1.473503}
  {\path{arXiv:https://doi.org/10.1063/1.473503}}, \href
  {http://dx.doi.org/10.1063/1.473503} {\path{doi:10.1063/1.473503}}.
\newline\urlprefix\url{https://doi.org/10.1063/1.473503}

\bibitem{Voter-hyperdyn-PRL1997}
A.~F. Voter,
  \href{https://link.aps.org/doi/10.1103/PhysRevLett.78.3908}{Hyperdynamics:
  Accelerated molecular dynamics of infrequent events}, Phys. Rev. Lett. 78
  (1997) 3908--3911.
\newblock \href {http://dx.doi.org/10.1103/PhysRevLett.78.3908}
  {\path{doi:10.1103/PhysRevLett.78.3908}}.
\newline\urlprefix\url{https://link.aps.org/doi/10.1103/PhysRevLett.78.3908}

\bibitem{Laio01102002}
A.~Laio, M.~Parrinello,
  \href{http://www.pnas.org/content/99/20/12562.abstract}{Escaping free-energy
  minima}, Proc. Natl. Acad. Sci. USA 99~(20) (2002) 12562--12566.
\newblock \href
  {http://arxiv.org/abs/http://www.pnas.org/content/99/20/12562.full.pdf}
  {\path{arXiv:http://www.pnas.org/content/99/20/12562.full.pdf}}, \href
  {http://dx.doi.org/10.1073/pnas.202427399}
  {\path{doi:10.1073/pnas.202427399}}.
\newline\urlprefix\url{http://www.pnas.org/content/99/20/12562.abstract}

\bibitem{Darve-ABF-JCP2001}
E.~Darve, A.~Pohorille, \href{https://doi.org/10.1063/1.1410978}{Calculating
  free energies using average force}, J. Chem. Phys. 115~(20) (2001)
  9169--9183.
\newblock \href {http://arxiv.org/abs/https://doi.org/10.1063/1.1410978}
  {\path{arXiv:https://doi.org/10.1063/1.1410978}}, \href
  {http://dx.doi.org/10.1063/1.1410978} {\path{doi:10.1063/1.1410978}}.
\newline\urlprefix\url{https://doi.org/10.1063/1.1410978}

\bibitem{Maeda-Morokuma-AFIR2010}
S.~Maeda, K.~Morokuma, \href{https://doi.org/10.1063/1.3457903}{Communications:
  A systematic method for locating transition structures of a+b→x type
  reactions}, J. Chem. Phys. 132~(24) (2010) 241102.
\newblock \href {http://arxiv.org/abs/https://doi.org/10.1063/1.3457903}
  {\path{arXiv:https://doi.org/10.1063/1.3457903}}, \href
  {http://dx.doi.org/10.1063/1.3457903} {\path{doi:10.1063/1.3457903}}.
\newline\urlprefix\url{https://doi.org/10.1063/1.3457903}

\bibitem{Allen-tenWolde-FFS-PRL2005}
R.~J. Allen, P.~B. Warren, P.~R. ten Wolde,
  \href{https://link.aps.org/doi/10.1103/PhysRevLett.94.018104}{Sampling rare
  switching events in biochemical networks}, Phys. Rev. Lett. 94 (2005) 018104.
\newblock \href {http://dx.doi.org/10.1103/PhysRevLett.94.018104}
  {\path{doi:10.1103/PhysRevLett.94.018104}}.
\newline\urlprefix\url{https://link.aps.org/doi/10.1103/PhysRevLett.94.018104}

\bibitem{Harada-Kitao-PaCSMD-JCP2013}
R.~Harada, A.~Kitao, \href{https://doi.org/10.1063/1.4813023}{Parallel cascade
  selection molecular dynamics (pacs-md) to generate conformational transition
  pathway}, J. Chem. Phys. 139~(3) (2013) 035103.
\newblock \href {http://arxiv.org/abs/https://doi.org/10.1063/1.4813023}
  {\path{arXiv:https://doi.org/10.1063/1.4813023}}, \href
  {http://dx.doi.org/10.1063/1.4813023} {\path{doi:10.1063/1.4813023}}.
\newline\urlprefix\url{https://doi.org/10.1063/1.4813023}

\bibitem{Harada-Shigeta-SDS-JCTC2017}
R.~Harada, Y.~Shigeta,
  \href{https://doi.org/10.1021/acs.jctc.6b01112}{Efficient conformational
  search based on structural dissimilarity sampling: Applications for
  reproducing structural transitions of proteins}, J. Chem. Theor. Comput.
  13~(3) (2017) 1411--1423.
\newblock \href {http://arxiv.org/abs/https://doi.org/10.1021/acs.jctc.6b01112}
  {\path{arXiv:https://doi.org/10.1021/acs.jctc.6b01112}}, \href
  {http://dx.doi.org/10.1021/acs.jctc.6b01112}
  {\path{doi:10.1021/acs.jctc.6b01112}}.
\newline\urlprefix\url{https://doi.org/10.1021/acs.jctc.6b01112}

\bibitem{Harada-Shigeta-SDS-JCC2017}
R.~Harada, Y.~Shigeta, \href{http://dx.doi.org/10.1002/jcc.24837}{Structural
  dissimilarity sampling with dynamically self-guiding selection}, J. Comput.
  Chem. 38~(22) (2017) 1921--1929.
\newblock \href {http://dx.doi.org/10.1002/jcc.24837}
  {\path{doi:10.1002/jcc.24837}}.
\newline\urlprefix\url{http://dx.doi.org/10.1002/jcc.24837}

\bibitem{Akashi}
R.~Akashi, Y.~S. Nagornov,
  \href{https://doi.org/10.7566/JPSJ.87.063801}{Stochastic formalism for
  thermally driven distribution frontier: A nonempirical approach to the
  potential escape problem}, Journal of the Physical Society of Japan 87~(6)
  (2018) 063801.
\newblock \href {http://arxiv.org/abs/https://doi.org/10.7566/JPSJ.87.063801}
  {\path{arXiv:https://doi.org/10.7566/JPSJ.87.063801}}, \href
  {http://dx.doi.org/10.7566/JPSJ.87.063801}
  {\path{doi:10.7566/JPSJ.87.063801}}.
\newline\urlprefix\url{https://doi.org/10.7566/JPSJ.87.063801}

\bibitem{Hilderbrandt-NewtonRaphson}
R.~L. Hilderbrandt,
  \href{http://www.sciencedirect.com/science/article/pii/0097848577850080}{Application
  of newton-raphson optimization techniques in molecular mechanics
  calculations}, Comput. Chem. 1~(3) (1977) 179 -- 186.
\newblock \href
  {http://dx.doi.org/https://doi.org/10.1016/0097-8485(77)85008-0}
  {\path{doi:https://doi.org/10.1016/0097-8485(77)85008-0}}.
\newline\urlprefix\url{http://www.sciencedirect.com/science/article/pii/0097848577850080}

\bibitem{Cerjan-Miller-eigenvec-follow}
C.~J. Cerjan, W.~H. Miller, \href{https://doi.org/10.1063/1.442352}{On finding
  transition states}, J. Chem. Phys. 75~(6) (1981) 2800--2806.
\newblock \href {http://arxiv.org/abs/https://doi.org/10.1063/1.442352}
  {\path{arXiv:https://doi.org/10.1063/1.442352}}, \href
  {http://dx.doi.org/10.1063/1.442352} {\path{doi:10.1063/1.442352}}.
\newline\urlprefix\url{https://doi.org/10.1063/1.442352}

\bibitem{gentlest-ascent-Zhou2011}
W.~E, X.~Zhou, \href{http://stacks.iop.org/0951-7715/24/i=6/a=008}{The gentlest
  ascent dynamics}, Nonlinearity 24~(6) (2011) 1831.
\newline\urlprefix\url{http://stacks.iop.org/0951-7715/24/i=6/a=008}

\bibitem{Dimer1999}
G.~Henkelman, H.~Jonsson, \href{https://doi.org/10.1063/1.480097}{A dimer
  method for finding saddle points on high dimensional potential surfaces using
  only first derivatives}, The Journal of Chemical Physics 111~(15) (1999)
  7010--7022.
\newblock \href {http://arxiv.org/abs/https://doi.org/10.1063/1.480097}
  {\path{arXiv:https://doi.org/10.1063/1.480097}}, \href
  {http://dx.doi.org/10.1063/1.480097} {\path{doi:10.1063/1.480097}}.
\newline\urlprefix\url{https://doi.org/10.1063/1.480097}

\bibitem{Gardiner-book}
C.~Gardiner, Stochastic Methods: A Handbook for the Natural and Social Sciences
  (4th edition), Springer, 2009.

\bibitem{Giardina-review-JStatPhys2011}
C.~Giardina, J.~Kurchan, V.~Lecomte, J.~Tailleur,
  \href{https://doi.org/10.1007/s10955-011-0350-4}{Simulating rare events in
  dynamical processes}, J. Stat. Phys. 145~(4) (2011) 787--811.
\newblock \href {http://dx.doi.org/10.1007/s10955-011-0350-4}
  {\path{doi:10.1007/s10955-011-0350-4}}.
\newline\urlprefix\url{https://doi.org/10.1007/s10955-011-0350-4}

\bibitem{Suzuki-Trotter-1-AMS1959}
H.~F. Trotter, On the product of semi-groups of operators, Proc. Amer. Math.
  Soc. 10 (1959) 545.

\bibitem{Suzuki-Trotter-2-CMP1976}
M.~Suzuki, \href{https://doi.org/10.1007/BF01609348}{Generalized trotter's
  formula and systematic approximants of exponential operators and inner
  derivations with applications to many-body problems}, Commun. Math. Phys.
  51~(2) (1976) 183--190.
\newblock \href {http://dx.doi.org/10.1007/BF01609348}
  {\path{doi:10.1007/BF01609348}}.
\newline\urlprefix\url{https://doi.org/10.1007/BF01609348}

\bibitem{note-Trotter}
Using the generalized Trotter formula [Refs. 28 and 29], one can implement the
  algorithms with error of arbitrary order with respect to $\tau$. Here, we
  have adopt the formula with error $O(\tau^3)$ for compromise between
  simplicity and accuracy for discrete time evolution.

\bibitem{Grassberger-importance-sampling-PRE1997}
P.~Grassberger,
  \href{https://link.aps.org/doi/10.1103/PhysRevE.56.3682}{Pruned-enriched
  rosenbluth method: Simulations of $\ensuremath{\theta}$ polymers of chain
  length up to 1 000 000}, Phys. Rev. E 56 (1997) 3682--3693.
\newblock \href {http://dx.doi.org/10.1103/PhysRevE.56.3682}
  {\path{doi:10.1103/PhysRevE.56.3682}}.
\newline\urlprefix\url{https://link.aps.org/doi/10.1103/PhysRevE.56.3682}

\bibitem{conversation_walkers}
T.~Brewer, S.~R. Clark, R.~Bradford, R.~L. Jack,
  \href{http://stacks.iop.org/1742-5468/2018/i=5/a=053204}{Efficient
  characterisation of large deviations using population dynamics}, Journal of
  Statistical Mechanics: Theory and Experiment 2018~(5) (2018) 053204.
\newline\urlprefix\url{http://stacks.iop.org/1742-5468/2018/i=5/a=053204}

\bibitem{LJ1990}
D.~J. Wales, R.~S. Berry, \href{https://doi.org/10.1063/1.457788}{Melting and
  freezing of small argon clusters}, The Journal of Chemical Physics 92~(7)
  (1990) 4283--4295.
\newblock \href {http://arxiv.org/abs/https://doi.org/10.1063/1.457788}
  {\path{arXiv:https://doi.org/10.1063/1.457788}}, \href
  {http://dx.doi.org/10.1063/1.457788} {\path{doi:10.1063/1.457788}}.
\newline\urlprefix\url{https://doi.org/10.1063/1.457788}

\bibitem{LJ1999_38}
J.~P.~K. Doye, M.~A. Miller, D.~J. Wales,
  \href{https://doi.org/10.1063/1.478595}{The double-funnel energy landscape of
  the 38-atom lennard-jones cluster}, The Journal of Chemical Physics 110~(14)
  (1999) 6896--6906.
\newblock \href {http://arxiv.org/abs/https://doi.org/10.1063/1.478595}
  {\path{arXiv:https://doi.org/10.1063/1.478595}}, \href
  {http://dx.doi.org/10.1063/1.478595} {\path{doi:10.1063/1.478595}}.
\newline\urlprefix\url{https://doi.org/10.1063/1.478595}

\bibitem{LJ1999_size}
J.~P.~K. Doye, M.~A. Miller, D.~J. Wales,
  \href{https://doi.org/10.1063/1.480217}{Evolution of the potential energy
  surface with size for lennard-jones clusters}, The Journal of Chemical
  Physics 111~(18) (1999) 8417--8428.
\newblock \href {http://arxiv.org/abs/https://doi.org/10.1063/1.480217}
  {\path{arXiv:https://doi.org/10.1063/1.480217}}, \href
  {http://dx.doi.org/10.1063/1.480217} {\path{doi:10.1063/1.480217}}.
\newline\urlprefix\url{https://doi.org/10.1063/1.480217}

\bibitem{LJ2000_38}
F.~Calvo, J.~P. Neirotti, D.~L. Freeman, J.~D. Doll,
  \href{https://doi.org/10.1063/1.481672}{Phase changes in 38-atom
  lennard-jones clusters. ii. a parallel tempering study of equilibrium and
  dynamic properties in the molecular dynamics and microcanonical ensembles},
  The Journal of Chemical Physics 112~(23) (2000) 10350--10357.
\newblock \href {http://arxiv.org/abs/https://doi.org/10.1063/1.481672}
  {\path{arXiv:https://doi.org/10.1063/1.481672}}, \href
  {http://dx.doi.org/10.1063/1.481672} {\path{doi:10.1063/1.481672}}.
\newline\urlprefix\url{https://doi.org/10.1063/1.481672}

\bibitem{LJ2006}
G.~Adjanor, M.~Ath{\`e}nes, F.~Calvo,
  \href{https://doi.org/10.1140/epjb/e2006-00353-0}{Free energy landscape from
  path-sampling: application to thestructural transition in lj38}, The European
  Physical Journal B - Condensed Matter and Complex Systems 53~(1) (2006)
  47--60.
\newblock \href {http://dx.doi.org/10.1140/epjb/e2006-00353-0}
  {\path{doi:10.1140/epjb/e2006-00353-0}}.
\newline\urlprefix\url{https://doi.org/10.1140/epjb/e2006-00353-0}

\bibitem{LJ2018_38}
J.~D. Mallory, V.~A. Mandelshtam,
  \href{https://doi.org/10.1063/1.5050410}{Quantum-induced solid-solid
  transitions and melting in the lennard-jones lj38 cluster}, The Journal of
  Chemical Physics 149~(10) (2018) 104305.
\newblock \href {http://arxiv.org/abs/https://doi.org/10.1063/1.5050410}
  {\path{arXiv:https://doi.org/10.1063/1.5050410}}, \href
  {http://dx.doi.org/10.1063/1.5050410} {\path{doi:10.1063/1.5050410}}.
\newline\urlprefix\url{https://doi.org/10.1063/1.5050410}

\end{thebibliography}

\end{document}